\documentclass[10.5pt,a4paper]{article}

\usepackage[margin=1in]{geometry}
\usepackage{setspace}
\usepackage{graphicx}
\usepackage{booktabs}
\usepackage{longtable}
\usepackage{amsmath}
\usepackage{amssymb}
\usepackage{natbib}
\usepackage{hyperref}
\usepackage{caption}
\usepackage{subcaption}
\usepackage{authblk}
\usepackage{appendix}
\usepackage{enumitem}
\usepackage{booktabs}
\usepackage{siunitx}
\usepackage{threeparttable}
\usepackage{array}
\usepackage{booktabs}
\usepackage{longtable}
\usepackage{tabularx}
\usepackage{threeparttable}
\usepackage{ragged2e}
\usepackage{array}
\newcolumntype{L}[1]{>{\raggedright\arraybackslash}p{#1}}
\usepackage[utf8]{inputenc}
\usepackage{multicol} 

\hypersetup{
    colorlinks=true,
    linkcolor=blue,
    citecolor=cyan,
    urlcolor=red
}

\title{\textbf{From Timbuktu to SKA: Who owns the Astronomy knowledge Africa produces?}}

\author[1,2]{Rhea Koch}
\author[3,4]{Amare Abebe}

\affil[1]{Centre for Higher Education Professional Development (CHEPD), North-West University, Potchefstroom 2520, South Africa}
\affil[2]{The Education and Human Rights in Diversity Research Unit (Edu-HRight), Potchefstroom 2520, South Africa}
\affil[3]{Centre for Space Research, North-West University, Potchefstroom 2520, South Africa}
\affil[4]{National Institute for Theoretical and Computational Sciences (NITheCS), Potchefstroom 2520, South Africa}
\date{}

\begin{document}

\maketitle

\begin{abstract}
Africa has deep and diverse traditions of astronomical knowledge, ranging from archaeological astronomy, pharaonic stellar timekeeping, and manuscript astronomy in Timbuktu and the Sahel, to ecological seasonal astronomy in southern Africa and calendrical computation in the Ethiopian tradition. These multiple epistemic traditions form part of a long intellectual history that precedes and intersects with contemporary astronomy on the continent.
In the context of this long-standing intellectual heritage, an important question arises regarding the contemporary circulation of African astronomical knowledge within the global scientific system. This study therefore examines where African astronomical knowledge is produced, validated, and circulated in modern scholarly communication. Using bibliometric data from the Web of Science Core Collection, our analysis investigates publication and citation patterns in the research area Astronomy and Astrophysics between 2000 and 2025, with particular attention to the publication venues of African scholars.
The findings highlight a critical paradox: while African researchers are active contributors to global astronomical discovery, the intellectual capital generated through this work is largely stored, validated, and circulated through publication systems located outside the continent. This pattern reflects broader asymmetries within global scholarly communication, where dominant publishing infrastructures shape visibility, citation impact, and authority.
By applying a decolonial lens to metrics such as citation impact and ownership, the paper calls for a critical reassessment of the academic practices that sustain epistemic coloniality. It concludes that achieving scientific equity requires a strategic shift in publication choices to build and fortify a sovereign African knowledge archive.
\end{abstract}

\vspace{0.5cm}

\newpage


\section{Introduction}

Humanity has contemplated the sky since time immemorial. Questions about origin, order, seasonality, destiny, and place in the cosmos are among the oldest recorded intellectual concerns. Long before the emergence of formal scientific institutions, communities across the globe developed systematic ways of observing, interpreting, and encoding celestial phenomena. In this sense, astronomy is as old as organised human society itself.

Africa occupies a singular position in this long tradition of \emph{doing astronomy}. As the cradle of humankind, the continent is not only central to biological origins but also to some of the earliest documented engagements with the sky. Archaeological evidence from sites such as Nabta Playa \citep{Malville1998,Malville2003, Wendorf2001}, along with later textual traditions preserved in manuscript cultures including Timbuktu \citep{Medupe2008,Jeppie2008,Hunwick2003}, and ethnographic records of San and Zulu cosmologies documenting detailed knowledge of prominent stars and asterisms \citep{Bleek1911,Hollmann2004,Berglund1976,Krige1950,Ruggles2015} all attest to sophisticated astronomical reasoning embedded in social, agricultural, and cosmological systems. Astronomy, in Africa as elsewhere, has always been a science of place.

In the contemporary era, astronomy remains profoundly tied to geography, but in new forms. From horizon calendars and stone alignments to modern interferometric arrays, astronomical knowledge has been shaped by where observations are made, how they are recorded, and who controls their interpretation. Today, astronomy is also a paradigmatic example of infrastructure-intensive, data-driven science, in which epistemic authority is inseparable from access to telescopes, archives, analysis pipelines, high-performance computing, and the validation mechanisms of scholarly communication.

Africa’s position within the global astronomy community has changed drastically over the past two decades, e.g, the number of professional astronomers as well as the share of peer-reviewed publications to the global astronomy knowledge has increased dramatically as shown in the figures below. 

\begin{figure}[!htb]
\centering
 \fbox{ \includegraphics[scale=0.55]{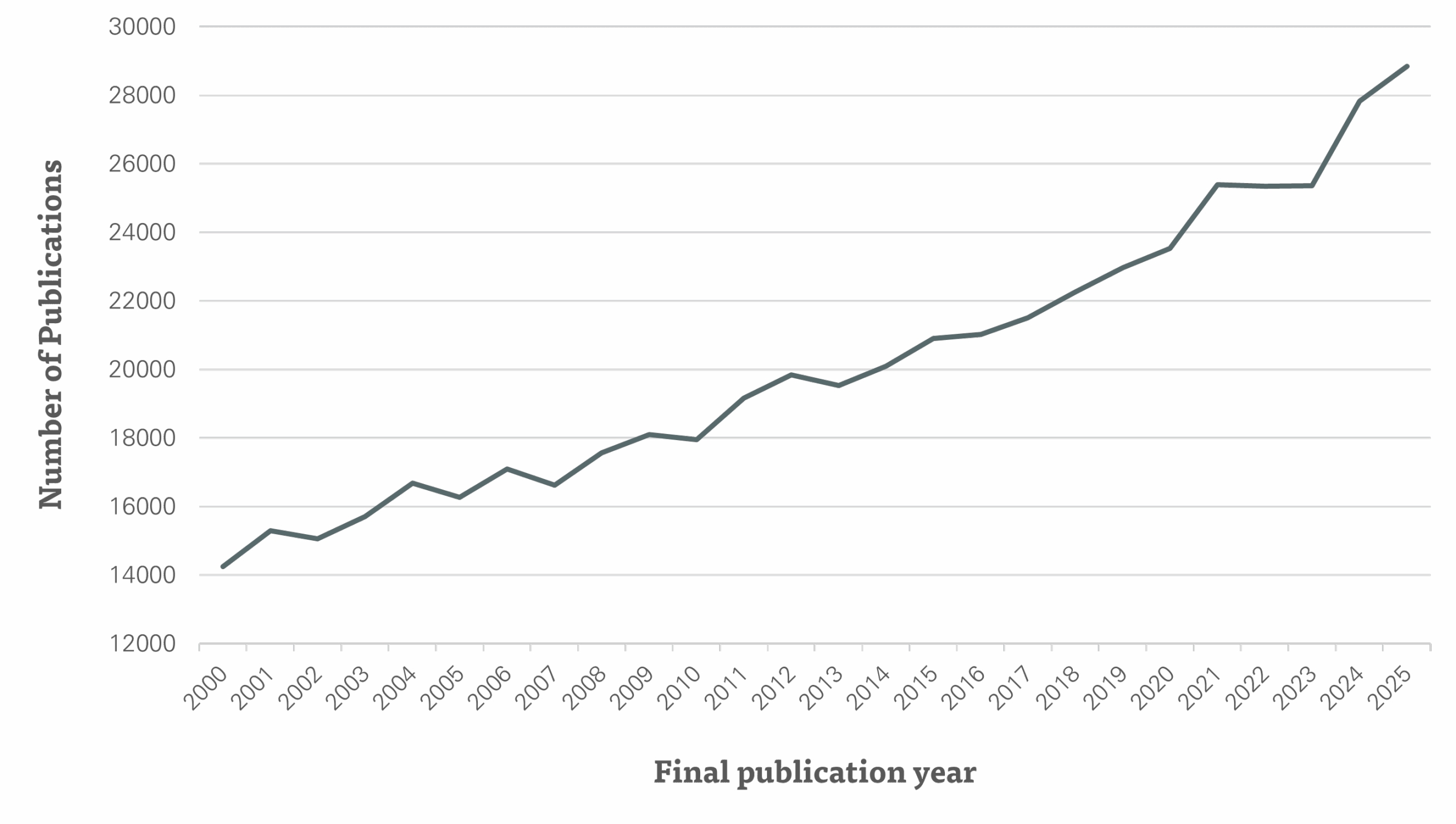}}
  \caption{Increasing number of African-affiliated publications in Astronomy and Astrophysics (as extracted from the Web of Science Core Collection (2020-2025).}
\end{figure}

 The commissioning of MeerKAT, the continued operation of SALT, and the hosting of major components of the SKA have placed the continent at the centre of global observational capacity. These developments are frequently cited as evidence of Africa’s integration into frontier science. While this integration is real and scientifically significant, it does not by itself resolve questions of authority, ownership, and long-term benefit.

This paper asks a deceptively simple question: where does African astronomy knowledge reside? We define \emph{knowledge residence} as the set of material, institutional, and epistemic locations through which knowledge is made legitimate, authoritative, and durable. It transcends the narrow understanding of knowledge production as equivalent to authorship or geographic participation. A paper may be written by an African scientist, draw on data collected on African soil, and address questions prioritised by African institutions, yet still reside elsewhere in the systems that store it, validate it, and circulate it. We distinguish three interrelated dimensions of knowledge residence:
\begin{itemize}
    \item  Material residence refers to the physical and digital infrastructures in which research objects are stored. This includes observational data, calibrated datasets, software pipelines, articles, preprints, metadata, repositories, and long-term archives. 
\item Institutional residence refers to the organisational structures through which legitimacy is conferred such as journals, publishers, indexers, funding agencies, peer review systems, university rankings, and research assessment frameworks.
\item Epistemic residence refers to interpretive authority: who defines the important questions, frames the meaning of findings, and determines the criteria by which knowledge is recognised as significant. This is the deepest layer of the framework.
\end{itemize}

This study draws on decolonial scholarship that argues that modern knowledge systems are historically structured by unequal distributions of authority \citep{mignolo2002,mignolo2003,ndlovu,salgado}.

Scientific participation does not automatically imply epistemic sovereignty. Scholars and institutions in Africa may contribute substantially to globally significant research while remaining dependent on external infrastructures for validation, circulation, and preservation. So the question is not whether African scientists participate in global collaborations - they do - but where the durable record of knowledge is stored, who validates it, and who is empowered by its circulation.

The rest of the paper is organised as follows. Section \ref{trad} outlines Africa’s astronomical knowledge traditions in historical perspective.  Section \ref{data} presents the data and methods, including the bibliometric dataset, query definitions, and citation-based indicators used in the study. We show our findings in Section \ref{results}, and then Section \ref{disc} discusses these findings in relation to the geopolitics of knowledge production, epistemic dependency, and the question of knowledge sovereignty in African astronomy. Section \ref{limit} considers the limitations of the study and offers recommendations for future research. Finally, Section \ref{conc} concludes the paper.

\section{Africa’s Astronomical Knowledge Traditions}\label{trad}
Africa’s astronomical traditions span archaeological, textual, and ethnographic forms, each representing distinct modes of systematic engagement with celestial regularities. These traditions differ in medium, institutional structure, and epistemic framing, yet all demonstrate that sustained observation of the sky has long been embedded in African intellectual life.

\subsection{Archaeological Astronomy: Nabta Playa and Early Horizon Alignments}

One of the earliest documented examples of structured astronomical engagement in Africa is found at Nabta Playa in southern Egypt. Excavations conducted by \cite{Wendorf2001} revealed megalithic stone alignments dating to the sixth millennium BCE, interpreted as having calendrical or stellar significance.  \cite{Malville1998} argue that certain alignments correspond to the rising positions of prominent stars and solar solstitial points, suggesting deliberate horizon-based astronomical orientation.

While the precise degree of intentional astronomical design remains debated, the site provides clear evidence that Neolithic pastoral communities in the Sahara engaged in systematic observation of seasonal celestial cycles. Such cycles would have been critical for rainfall prediction, migration, and subsistence planning. Nabta Playa thus represents an early material instantiation of what may be termed ecological astronomy: the encoding of seasonal regularities in built landscape.

\subsection{Pharaonic Stellar Timekeeping and Decanal Astronomy}

In pharaonic Egypt, astronomical observation became institutionalised within temple and state contexts. Egyptian decanal star clocks, preserved in coffin texts and temple ceilings, divided the night into twelve segments using the heliacal rising of specific stars \citep{Neugebauer1969full,Clagett1995}. These ``decans'' functioned as stellar timekeepers, regulating ritual, labour cycles, and administrative coordination.

Astronomical ceilings, such as those found in the tomb of Senenmut and later in the Dendera zodiac, illustrate increasingly formalised stellar catalogues. These systems demonstrate not merely symbolic cosmology but operational astronomical timekeeping embedded within state structures. The integration of astronomical calculation with political and religious authority is already visible here: celestial order and terrestrial governance were conceptually linked.
\subsection{Manuscript Astronomy in Timbuktu and the Sahel}
Beyond archaeological alignments and oral cosmologies, West Africa preserves 
a substantial written tradition of mathematical astronomy. The manuscript 
collections of Timbuktu and surrounding Sahelian centres contain works on 
astronomical calculation, calendrical reckoning, planetary motion, and 
zodiacal divisions, reflecting engagement with the broader Islamic scientific 
tradition \citep{Hunwick2003,Jeppie2008,Brentjes2007,Medupe2008}.

Among these are manuscripts containing circular zodiacal diagrams 
(\textit{al-burūj}), tabulated numerical grids, and computational layouts 
consistent with the \textit{zīj} tradition of Islamic astronomy. The 
\textit{zīj} genre, developed from late antique Greek sources and refined 
through Persian and Arab mathematical astronomy, comprised astronomical 
handbooks containing planetary tables, trigonometric values, and calendrical 
algorithms used for timekeeping, ritual regulation, and astrological 
interpretation \citep{King2005,Saliba1994}. Copies and adaptations of such 
materials circulated widely across North and West Africa from the medieval 
period onward.

The presence of zodiacal wheels and tabulated numerical structures in 
Timbuktu manuscripts demonstrates not merely symbolic cosmology but 
computational engagement with celestial cycles. 
\begin{figure}[htbp]
\centering
\includegraphics[width=0.85\textwidth]{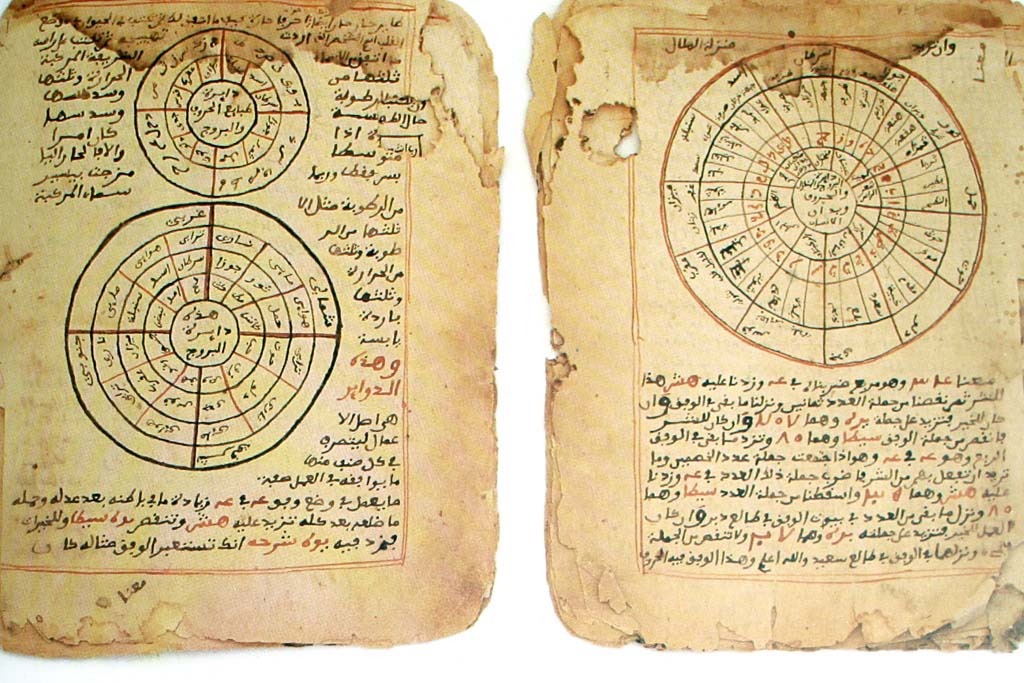}
\caption{Zodiacal diagram (\textit{al-burūj}) from a Timbuktu manuscript. 
The concentric circular structure represents the twelve zodiac signs divided 
into degrees, a format characteristic of the Islamic astronomical 
\textit{zīj} tradition. Such diagrams were used for calendrical computation, 
planetary position estimation, and astrological interpretation. 
The presence of this format in West African manuscripts demonstrates 
participation in trans-Saharan mathematical astronomy networks. \href{https://commons.wikimedia.org/w/index.php?curid=8654070}{[Credit: Elias Altmimi - EurAstro : Mission to Mali, Public Domain.]}}
\label{fig:zodiac}
\end{figure}

These texts situate West 
African scholarship within trans-Saharan intellectual networks linking 
Cairo, Fez, Andalusia, and the Sahel. As \cite{Medupe2008} argues, 
the astronomical materials preserved in Timbuktu reflect sustained 
scholarly participation in global mathematical astronomy rather than 
isolated local tradition.

These manuscript traditions are significant for two reasons. First, they 
demonstrate that Africa’s engagement with formal astronomical calculation 
predates modern observatory science on the continent. Second, they reveal 
that questions of authority and validation have long been mediated through 
manuscript circulation, commentary, and scholarly networks - an earlier 
form of the infrastructural dynamics examined in this paper.

\begin{figure}[htbp]
\centering
\includegraphics[width=0.78\textwidth]{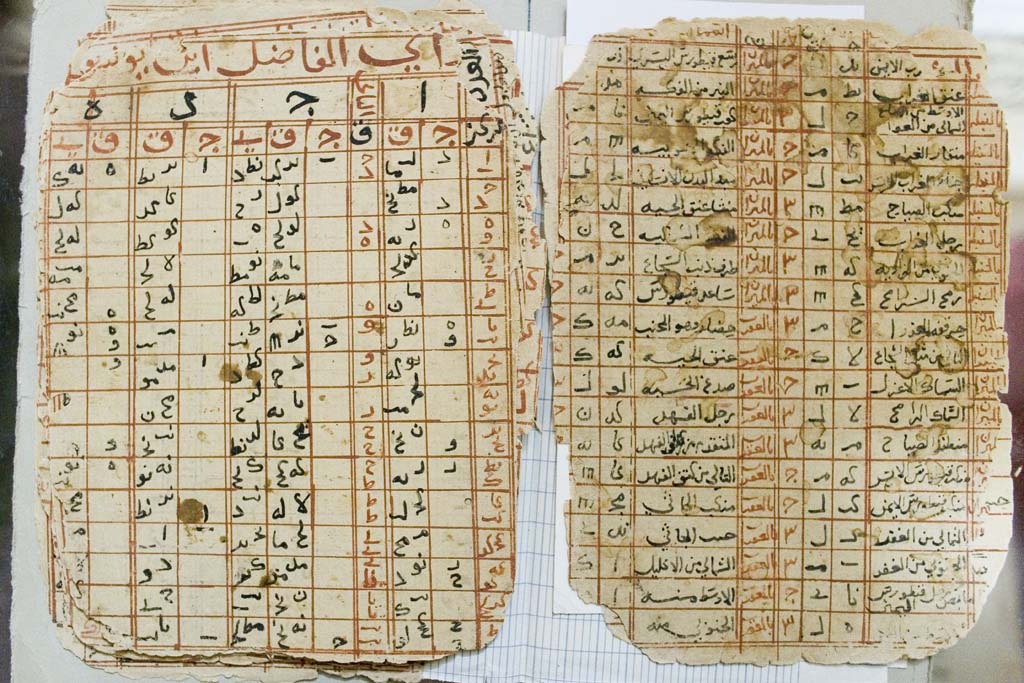}
\caption{Tabulated numerical grids from a Sahelian manuscript tradition. 
The structured layout of numerical columns and letter–number correspondences 
is consistent with astronomical and calendrical tables derived from the 
Islamic \textit{zīj} genre. Such tables were used for timekeeping, 
lunar calculations, planetary motion estimation, and ritual calendar 
determination. These materials situate West African scholarship within 
the broader computational astronomy tradition of the medieval Islamic world. \href{https://commons.wikimedia.org/w/index.php?curid=8654181?}{[Credit: Unknown author - EurAstro: Mission to Mali, Public Domain.]}}
\label{fig:tables}
\end{figure}

\subsection{Ecological and Seasonal Astronomy in Southern Africa}

In southern Africa, astronomical knowledge is preserved primarily through ethnographic records of San and Nguni-speaking communities. The Bleek and Lloyd archive documents detailed \textbar Xam associations of prominent stars and asterisms with seasonal transitions, subsistence cycles, and ritual timing \citep{Bleek1911,Hollmann2004}. The heliacal rising of the Pleiades, for example, has been linked to seasonal change and social activity in several southern African traditions.

Zulu cosmology similarly encodes stellar observation within agricultural and ritual frameworks. The Pleiades (\textit{isiLimela}) were associated with the onset of the planting season, while lunar cycles structured ritual calendars and social coordination \citep{Berglund1976,Krige1950}. These traditions reflect systematic empirical engagement with recurrent celestial phenomena, even in the absence of formalised written tables.

It is important to distinguish such ecological astronomy from long-distance navigational traditions found in other regions of the world. In southern Africa, celestial observation was primarily tied to seasonal regulation, social organisation, and ritual life rather than maritime navigation. Nevertheless, the epistemic structure remains clear: recurring astronomical regularities were identified, named, and integrated into predictive cultural systems.

\subsection{Calendrical Computation in Ethiopian Christian Tradition}

In the Horn of Africa, astronomical computation was embedded within ecclesiastical institutions. The Ethiopian calendar, derived from late antique Alexandrian computus traditions, required precise calculation of lunar cycles, solar years, and the date of Easter \citep{Mosshammer2008,Yaqob2007}. Ethiopian monastic scholars preserved computational tables in Ge’ez manuscripts, integrating solar and lunar reckoning into liturgical regulation and ecclesiastical governance \citep{Haile1981,Tamrat1972}. Manuscript traditions document the transmission of calendrical algorithms, epact calculations, and Paschal tables, reflecting continuity with broader Mediterranean computistical scholarship while adapting it within Ethiopian Christian institutional structures \citep{Neugebauer1975}.

This tradition represents another mode of institutional astronomy: celestial calculation embedded within religious authority structures. As in pharaonic Egypt and Timbuktu scholarship, astronomical knowledge functioned within organised textual systems that mediated authority through manuscript transmission and scholarly lineage.

\subsection{Multiple Epistemic Forms of African Astronomy}

Across the continent, African astronomical knowledge appears in multiple epistemic forms: 
megalithic alignment, temple inscription, manuscript computation, ecclesiastical calendrics, 
and oral seasonal encoding \citep{Ruggles2015,Neugebauer1975}. These are not isolated traditions 
but distinct instantiations of systematic engagement with celestial regularities.

What varies across these traditions is not the presence of structured astronomical reasoning, 
but the institutional form through which it is preserved and validated. In prehistoric contexts, 
authority was encoded in landscape and monument. In pharaonic Egypt, it was embedded in temple 
and state ritual \citep{Clagett1995}. In Timbuktu, it circulated through manuscript networks 
linked to trans-Saharan scholarly exchange \citep{Hunwick2003,Jeppie2008}. In Ethiopia, 
astronomical computation was institutionalised within monastic and ecclesiastical structures 
through calendrical tables and computistical manuscripts \citep{Yaqob2007,Haile1981}. In southern 
Africa, it was transmitted orally within social institutions regulating seasonal and ritual life 
\citep{Berglund1976,Hollmann2004}. Each represents a different infrastructure of knowledge residence.

Understanding these earlier infrastructures provides historical context for the contemporary 
question posed in this paper: 
when astronomy is institutionalised through observatories, 
archives, journals, and global collaborations, where does its epistemic authority ultimately reside?

\section{Bibliometric Data Analysis}\label{data}

Bibliometric data analysis is an important way to measure, monitor, and study research outputs.  It also provides a means to identify publication patterns \citep{Matorevhu2024}. Bibliometric indicators are used in this study not because they provide a complete account of scientific value, but because they reveal something important about knowledge residence. 

 These indicators help trace the geography of validation and visibility within contemporary astronomy, even while remaining themselves embedded in unequal systems of coverage and recognition.
 
Data were obtained from InCites Benchmarking \& Analytics (Clarivate), based on the Web of Science Core Collection. The dataset comprised all documents indexed under the Web of Science research area \emph{Astronomy \& Astrophysics} published between 2000 and 2025. 
 
African-affiliated documents were identified using institutional address metadata corresponding to African countries. A document was classified as {\it African-affiliated} if at least one author listed an institutional address within an African country. A document was classified as an {\it African-led publication} if the first author is using an African institution as their affiliation. 

Publication source location was determined using country-of-registration metadata provided within Web of Science. Journals and publishers head-quartered in an African country were classified as {\it African-based} sources.

Descriptive bibliometric analysis was conducted to quantify (i) Africa’s share of global document output, (ii) the proportion of African-based sources within the indexed publication set, (iii) the proportion of African-led publications within the indexed publication set and (iv) the proportion of African-affiliated documents published in African-based versus non-African-based sources

\subsection{Impact Indicators}

Because this paper asks where African astronomy knowledge \emph{resides}, we treat citation-based indicators not as intrinsic measures of scientific quality but as proxies for recognition within a specific validation infrastructure: indexed publications and citation tracking as implemented in Web of Science Core Collection and InCites \citep{InCitesCitationImpact2024,InCitesCNCI2025}. 

\paragraph{Citation Impact (CI).}
In InCites, Citation Impact is defined as total citations divided by total publications for a set of documents, yielding the average citations per document \citep{InCitesCitationImpact2024}:
\begin{equation}
CI(S) = \frac{\sum_{d \in S} C_d}{|S|},
\end{equation}
where $C_d$ is the citation count of document $d$ and $|S|$ is the number of documents in set $S$.
CI is transparent and intuitive but can be dominated by outliers and does not reflect the total volume of research output beyond averaging \citep{InCitesCitationImpact2024}.

\paragraph{Category Normalized Citation Impact (CNCI).}
To compare impact across publication years and categories, we use Category Normalized Citation Impact (CNCI), defined in InCites as the ratio of observed citations $C_d$ to the expected citation rate $E_d$ for documents of the same year, document type, and subject category \citep{InCitesCNCI2025}:
\begin{equation}
CNCI(d) = \frac{C_d}{E_d}.
\end{equation}
For multi-category assignments, InCites applies a harmonic-mean approach to reduce dominance by a single high-performing category \citep{InCitesCNCI2025}. For a document set, CNCI is reported as the arithmetic mean of document-level CNCI values \citep{InCitesCNCI2025}:
\begin{equation}
CNCI(S) = \frac{1}{|S|}\sum_{d \in S} CNCI(d).
\end{equation}
A CNCI of 1 indicates performance at the global baseline; values above 1 indicate above-average citation uptake relative to comparable papers \citep{InCitesCNCI2025}.

\paragraph{Why use both indicators?}
CI captures \emph{raw visibility} of African-affiliated outputs within indexed literature. CNCI captures \emph{relative recognition} after controlling for field/year/type effects, enabling fairer comparison across time and heterogeneous portfolios \citep{InCitesCNCI2025,Waltman2016Review}.

\paragraph{Interpreting metrics responsibly in African astronomy.}
Citation distributions are highly skewed, making averages sensitive to extreme values (especially for small subsets), and citation indicators inherit biases from database coverage and classification systems \citep{Seglen1992JIF,Mongeon2016Coverage,Asubiaro2024RegionalDisparities}. InCites itself cautions that CNCI can fluctuate strongly for recent years and small sets, recommending use alongside complementary indicators and broader time windows \citep{InCitesCNCI2025}. Following responsible metrics guidance, we therefore interpret CI and CNCI as components of an evidentiary bundle rather than standalone measures of value \citep{DORA2013,Hicks2015Leiden,Wilsdon2015MetricTide}.

\paragraph{Operationalising ``does impact vary by publication location?''}
In the analyses that follow, we report CI and CNCI for African-affiliated and African-led subsets, stratified by publication source location (African-based vs non-African-based journals), in order to assess whether citation visibility and relative recognition vary systematically with the infrastructures through which knowledge is circulated and validated.

\section{Results}\label{results}

The bibliometric results reveal several clear quantitative asymmetries in the production, leadership, and validation of African astronomy knowledge.

First, African participation in the global astronomy literature is substantial but remains a small fraction of total output. Of the 647,347 astronomy and astrophysics publications (Table~\ref{tab:global_africa_sa}) indexed in the Web of Science between 2000 and 2025, 16,023 (approximately 2.5\%) include at least one African-affiliated author (Table~\ref{tab:overview_totals}). This indicates marginal integration into global research activity for a continent that accounts for about 20\% of the global population.

\begin{table}[htbp]
\centering
\small
\caption{Global publication and citation metrics in Astronomy and Astrophysics (Web of Science), including Africa as a collective aggregate.}
\label{tab:global_africa_sa}

\setlength{\tabcolsep}{4pt}
\renewcommand{\arraystretch}{0.95}

\begin{tabular}{rlrrrr}
\toprule
\textbf{Rank} & \textbf{Name} & \textbf{Docs} & \textbf{Times cited} & \textbf{CI} & \textbf{CNCI} \\
\midrule

\multicolumn{2}{l}{Publications worldwide on WoS} & 647347 & 18586142 & 28.71 & 1.06 \\
\multicolumn{2}{l}{Publications worldwide on WoS with listed country} & 633556 & 18527786 & 29.24 & 1.07 \\
\midrule

1  & USA & 254706 & 10707210 & 42.04 & 1.37 \\
2  & Germany & 105894 & 4543832 & 42.91 & 1.45 \\
3  & United Kingdom & 98331 & 4343129 & 44.17 & 1.48 \\
4  & France & 73279 & 3120856 & 42.59 & 1.48 \\
5  & Italy & 71521 & 2687412 & 37.58 & 1.36 \\
6  & China mainland & 67428 & 1562532 & 23.17 & 0.98 \\
7  & Japan & 52875 & 1869568 & 35.36 & 1.23 \\
8  & Spain & 50941 & 1989680 & 39.06 & 1.46 \\
9  & Russia & 43649 & 1122914 & 25.73 & 0.81 \\
10 & Canada & 36924 & 1787646 & 48.41 & 1.67 \\
11 & Netherlands & 33490 & 1636190 & 48.86 & 1.78 \\
12 & India & 32636 & 808958 & 24.79 & 0.97 \\
13 & Australia & 30352 & 1375024 & 45.30 & 1.66 \\
14 & Switzerland & 26162 & 1337623 & 51.13 & 1.88 \\
15 & Chile & 24848 & 1032048 & 41.53 & 1.54 \\
16 & Brazil & 21084 & 659059 & 31.26 & 1.12 \\
17 & Poland & 19724 & 724857 & 36.75 & 1.28 \\
18 & South Korea & 18135 & 578333 & 31.89 & 1.15 \\
19 & Sweden & 17497 & 782795 & 44.74 & 1.61 \\

20 & \textbf{Africa (total)} & \textbf{16023} & \textbf{596102} & \textbf{37.20} & \textbf{1.52} \\

21 & Belgium & 15890 & 627257 & 39.47 & 1.46 \\
22 & Mexico & 14984 & 449502 & 30.00 & 1.10 \\
23 & Taiwan & 12312 & 470061 & 38.18 & 1.43 \\
24 & Denmark & 12066 & 650571 & 53.92 & 2.15 \\
25 & Austria & 11288 & 448531 & 39.74 & 1.42 \\
26 & \textbf{South Africa} & \textbf{11052} & \textbf{489114} & \textbf{44.26} & \textbf{1.66} \\

\bottomrule
\end{tabular}
\end{table}
However, this participation does not translate proportionally into leadership. When restricting to first-author affiliation as a proxy for intellectual leadership, the number drops to 5,713 publications, corresponding to only 35.7\% of African-affiliated outputs. Put differently, nearly two-thirds of Africa’s contributions occur within collaborations led externally.

\begin{figure}[htbp]
\centering
\fbox{\includegraphics[width=0.8\textwidth]{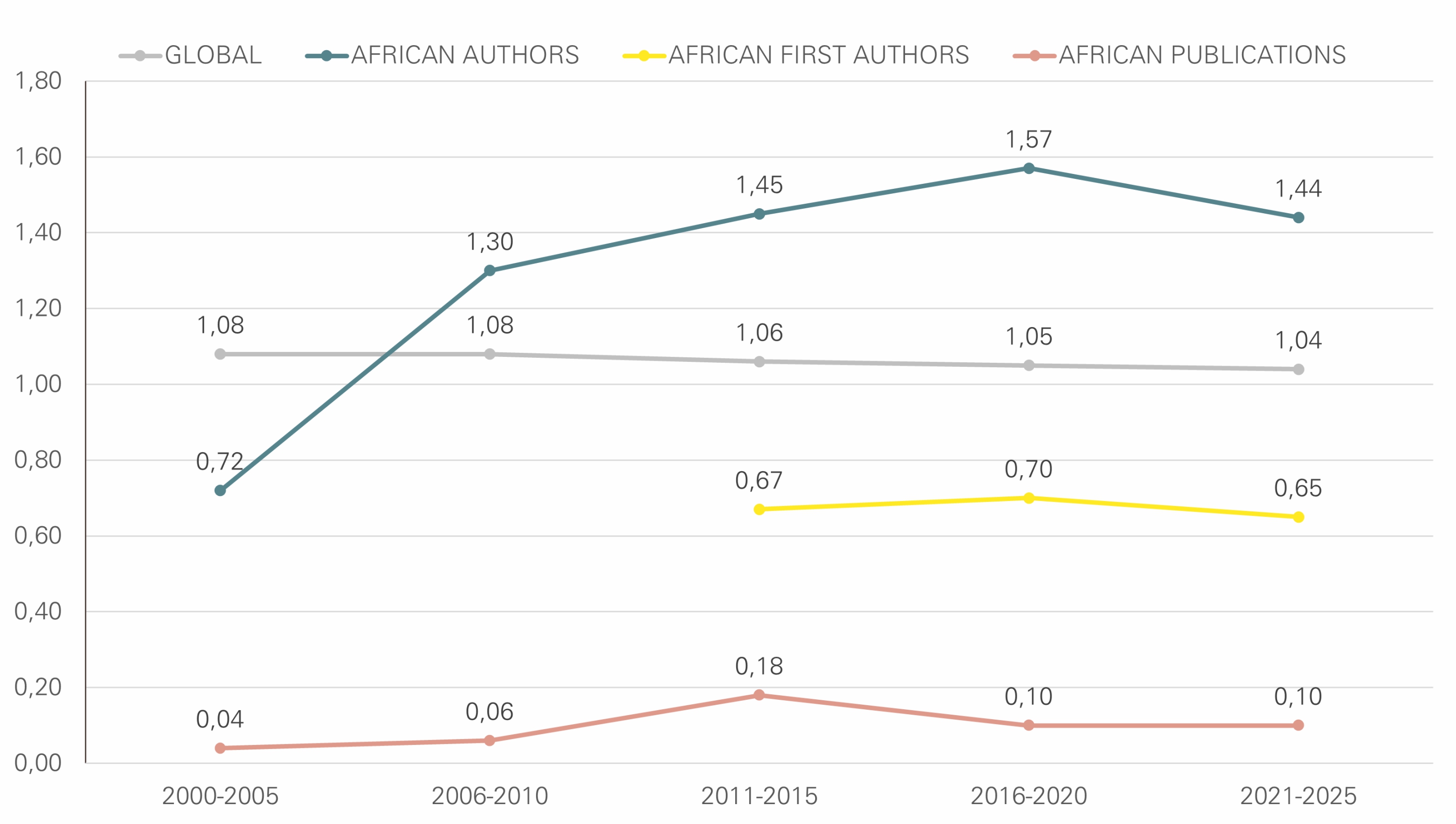}}
\caption{Temporal evolution of CNCI for global publications, African-affiliated publications, and African-led publications.}
\label{fig:cnci_trends}
\end{figure}

\begin{table}[htbp]
\centering
\small
\caption{Summary publication and citation indicators for the Web of Science dataset and African-affiliation subsets.}
\label{tab:overview_totals}
\begin{threeparttable}
\begin{tabularx}{\textwidth}{>{\raggedright\arraybackslash}p{5.2cm}rrrr}
\toprule
\textbf{Category} & \textbf{Docs} & \textbf{Times cited} & \textbf{CI} & \textbf{CNCI} \\
\midrule
Global & 647347 & 18586142 & 28.71 & 1.06 \\
African-affiliated author & 16023 & 596102 & 37.20 & 1.43 \\
First author African-affiliated & 5713 & 93961 & 16.45 & 0.69 \\
Published in Africa-based sources & 58 & 269 & 4.64 & 0.10 \\
African affiliated first author in Africa-based publication & 39 & 153 & 3.92 & 0.09 \\
African affiliated author in Africa-based publication & 19 & 109 & 5.74 & 0.14 \\
\bottomrule
\end{tabularx}

\begin{tablenotes}
\footnotesize
\item Docs = Web of Science documents; CI = Citation Impact; CNCI = Category Normalized Citation Impact.
\end{tablenotes}
\end{threeparttable}
\end{table}
Second, citation performance diverges sharply between participation and leadership as illustrated in (Figure~\ref{fig:cnci_trends}). African-affiliated publications achieve a Citation Impact (CI) of 37.20 and a Category Normalized Citation Impact (CNCI) of 1.43, both above the global benchmark values (CI = 28.71; CNCI = 1.06). This indicates that African-linked research is highly visible within the indexed citation network. By contrast, African-led publications exhibit substantially lower impact (CI = 16.45; CNCI = 0.69), falling below the world average. The CNCI for African-affiliated papers is therefore about 35\% above the global baseline, whereas the CNCI for African-led papers is about 35\% below it. This gap quantitatively captures an affiliation--leadership asymmetry in recognition.

Third, the location of publication venues shows an extreme concentration outside the continent.

\begin{table}[htbp]
\centering
\small
\caption{Publication and citation metrics for selected Africa publication sources in the Web of Science dataset.}
\label{tab:africa_publication_sources}
\begin{threeparttable}
\begin{tabularx}{\textwidth}{>{\RaggedRight\arraybackslash}p{6.0cm}r>{\RaggedRight\arraybackslash}p{2.8cm}rr}
\toprule
\textbf{Name} & \textbf{Docs} & \textbf{Source country/region} & \textbf{CI} & \textbf{CNCI} \\
\midrule
South African Journal of Science & 29 & South Africa & 2.07 & 0.05 \\
Scientific African & 3 & Rwanda & 2.33 & 0.11 \\
International Journal of the Physical Sciences & 4 & Nigeria & 3.50 & 0.07 \\
Momona Ethiopian Journal of Science & 2 & Ethiopia & 1.50 & 0.05 \\
Journal of Advanced Research & 16 & Egypt & 11.25 & 0.23 \\
Journal of Fundamental and Applied Sciences & 4 & Algeria & 1.25 & 0.04 \\
\bottomrule
\end{tabularx}
\begin{tablenotes}
\footnotesize
\item Docs = Web of Science documents; CI = citation impact; CNCI = category normalised citation impact.
\end{tablenotes}
\end{threeparttable}
\end{table}

Only 58 publications in the full dataset were published in Africa-based journals, representing less than 0.01\% of global output and only about 0.36\% of African-affiliated output (Table~\ref{tab:africa_publication_sources}). These publications also exhibit very low citation performance (CI = 4.64; CNCI = 0.10), far below both the global benchmark and the African-affiliated subset. Within this already small set, African-affiliated first-author papers published in Africa-based journals number only 39, with CI = 3.92 and CNCI = 0.09, while African-affiliated papers in Africa-based journals more broadly number just 19, with CI = 5.74 and CNCI = 0.14. The indexed validation infrastructure therefore overwhelmingly privileges non-African publication channels.

\begin{table}[htbp]
\centering
\small
\caption{Top 10 African countries in Astronomy and Astrophysics by publication output (Web of Science dataset).}
\label{tab:top10_countries}
\begin{tabular}{lrrrr}
\toprule
\textbf{Country} & \textbf{Docs} & \textbf{Cited} & \textbf{CNCI} \\
\midrule

South Africa & 11052 & 489114 & 1.66 \\
Egypt & 2176 & 49567 & 0.79 \\
Morocco & 1151 & 47287  & 1.57 \\
Nigeria & 689 & 8425 & 0.57 \\
Algeria & 670 & 30595 & \textbf{2.38} \\
Ethiopia & 327 & 4087 & 0.62 \\
Namibia & 283 & 22669 & \textbf{2.57} \\
Tunisia & 225 & 2162 & 0.39 \\
Uganda & 108 & 1763 & 0.80 \\
Benin & 100 & 2638 & 0.87 \\

\bottomrule
\end{tabular}
\end{table}
Fourth, the country-level and collaboration patterns reinforce this asymmetry. South Africa is by far the dominant continental actor, with 11,052 publications and 489,114 citations, yielding CI = 44.26 and CNCI = 1.66 (Table~\ref{tab:top10_countries}). Morocco follows with 1,151 publications (CNCI = 1.57), whereas Egypt records 2,176 publications but a markedly lower CNCI of 0.79. At the institutional level, the most productive African organisations are concentrated in South Africa and a small number of North African systems. The University of Cape Town alone contributes 3,016 publications, the National Research Foundation of South Africa 2,996, and the South African Astronomical Observatory 2,368 (Table~\ref{tab:wos_institution_metrics}). Several South African institutions also display particularly strong citation performance, including the University of KwaZulu-Natal (CI = 77.74, CNCI = 3.01), Stellenbosch University (CI = 61.85, CNCI = 3.08), and the University of Pretoria (CI = 82.76, CNCI = 4.53).

\begin{table}[htbp]
\centering
\small
\caption{Top 10 African institutions in Astronomy and Astrophysics by publication output (Web of Science dataset).}
\label{tab:wos_institution_metrics}
\begin{tabular}{L{5cm}rrrr}
\toprule
\textbf{Institution} & \textbf{Docs} & \textbf{Cited} & \textbf{Impact} & \textbf{CNCI} \\
\midrule

University of Cape Town & 3016 & 135914 & 45.06 & 1.57 \\
National Research Foundation (SA) & 2996 & 134539 & 44.91 & 1.52 \\
South African Astronomical Observatory & 2368 & 110982 & 46.87 & 1.58 \\
Egyptian Knowledge Bank & 2125 & 48233 & 22.70 & 0.79 \\
University of the Western Cape & 1683 & 87024 & 51.71 & 2.23 \\
North-West University & 1400 & 49574 & 35.41 & 1.33 \\
University of KwaZulu-Natal & 1365 & 106117 & 77.74 & 3.01 \\
University of the Witwatersrand & 1216 & 57809 & 47.54 & 1.91 \\
Rhodes University & 1164 & 56739 & 48.74 & 2.09 \\
Morocco (Cadi Ayyad University) & 743 & 34968 & 47.06 & 1.69 \\

\bottomrule
\end{tabular}
\end{table}
Finally, the highest citation impacts in Africa-linked output tend to cluster around collaboration with globally central research systems. In the collaboration table, partnerships involving the United States (CNCI = 2.38), the United Kingdom (CNCI = 2.55), Germany (CNCI = 2.70), France (CNCI = 2.83), the Netherlands (CNCI = 3.40), and Switzerland (CNCI = 3.70) all perform well above the global baseline (Table~\ref{tab:africa_collaboration_top25}). By contrast, several African collaboration partners remain below or only near the world benchmark, for example Egypt (CNCI = 0.96), Nigeria (CNCI = 0.73), Ethiopia (CNCI = 0.75), Botswana (CNCI = 0.66), and Ghana (CNCI = 0.47), although South Africa itself stands above baseline at 1.82 and Namibia reaches 2.59. Collaboration thus appears to function simultaneously as a visibility accelerator and as a channel through which recognition remains tied to externally dominant systems.
\begin{table}[htbp]
\centering
\caption{Top collaborating countries with African astronomy, ranked by number of co-authored publications.}
\label{tab:africa_collaboration_top25}

\setlength{\tabcolsep}{4pt}
\renewcommand{\arraystretch}{0.95}

\begin{tabular}{>{\RaggedRight\arraybackslash}p{3.2cm}rrrr}
\toprule
\textbf{Name} & \textbf{Docs} & \textbf{Times cited} & \textbf{CI} & \textbf{CNCI} \\
\midrule

Baseline for all items & 12846 & 555743 & 43.26 & 1.76 \\

\midrule

South Africa & 9489 & 462687 & 48.76 & 1.82 \\
USA & 6543 & 416957 & 63.73 & 2.38 \\
United Kingdom & 5709 & 390573 & 68.41 & 2.55 \\
Germany & 4773 & 348036 & 72.92 & 2.70 \\
France & 4151 & 308954 & 74.43 & 2.83 \\
Italy & 3886 & 267472 & 68.83 & 2.78 \\
Spain & 3170 & 257408 & 81.20 & 3.12 \\
Australia & 3125 & 221506 & 70.88 & 2.64 \\
Netherlands & 2630 & 223901 & 85.13 & 3.40 \\
China mainland & 2526 & 165421 & 65.49 & 2.76 \\
Poland & 2296 & 193658 & 84.35 & 2.99 \\
India & 2281 & 161654 & 70.87 & 2.69 \\
Chile & 2196 & 185862 & 84.64 & 3.36 \\
Canada & 2117 & 192916 & 91.13 & 3.48 \\
Switzerland & 2050 & 204214 & 99.62 & 3.70 \\
Japan & 2001 & 149996 & 74.96 & 2.99 \\
Russia & 1904 & 154572 & 81.18 & 2.67 \\
Brazil & 1889 & 141365 & 74.84 & 2.83 \\
Austria & 1648 & 120594 & 73.18 & 2.65 \\
Sweden & 1572 & 157408 & 100.13 & 4.04 \\
Egypt & 1557 & 43707 & 28.07 & 0.96 \\

\bottomrule
\end{tabular}
\end{table}
Taken together, our analysis shows that:
\begin{itemize}
    \item Africa’s knowledge production is about ten times less than our contribution to the global population. 
\item African astrophysical research is predominantly published in external journals and platforms;
\item High-impact publication, indexing, and citation systems are largely located outside the continent;
\item Despite African researchers’ citation impact being above the global baseline, research leadership (as proxied by first authorship) is well below this baseline.
\end{itemize}
As a result, African scholarship is integrated into global systems on externally defined terms.

\section{Discussion}\label{disc}
What emerges from the findings is not only disparity, but a patterned geography of knowledge that calls for engagement with the geopolitics of knowledge. The {\it geopolitics of knowledge} refers to the power-laden relationship between geography and the production, validation, and circulation of scientific thought. Based on the work of decolonial scholars such as \cite{mignolo2002,mignolo2003} and Quijano \citep{salgado}, the concept asserts that knowledge is never universal or neutral, but always located within specific geopolitical and historical power structures.

In the African context, scholars like \cite{ndlovu} have used this lens to examine \emph{epistemic coloniality}, referring to the process by which the authority to validate truth remains concentrated in external centres of power despite political independence. In a field like astronomy, where large-scale international collaborations and global publication systems shape research visibility, these dynamics raise important questions about how knowledge produced in Africa enters and circulates within the global system.

\subsection{Hegemony of Global Knowledge Production}
The global system of knowledge production has been described as hegemonic in the sense that dominant scientific traditions establish the interpretive frameworks that define what counts as legitimate knowledge. Within such a system, alternative perspectives or knowledge systems often struggle to gain recognition within mainstream academic structures \citep{cloete}. This hegemony is not only discursive, but institutional and infrastructural, shaping where knowledge is produced, validated, and ultimately recognised as authoritative \citep{gebremariam}.

Within this system, publication venues and citation infrastructures play a central role in locating epistemic authority. The designation of certain journals as international functions as a proxy for quality and legitimacy, positioning them as default sites of validation. As a result, knowledge becomes authoritative not only through its content, but through where it is published and how it circulates within global indexing systems.

The bibliometric finding that African astronomy is predominantly published outside the continent (\emph{international}) reflects this broader configuration of epistemic authority. When the international journal becomes the assumed end point of high-quality research, it reinforces a geography in which Africa is positioned as a site of scientific labour, but not as a site of scientific authority. Scholars like \cite{mungwini} and \cite{cloete} argue that such patterns sustain a colonial value system that continues to reproduce perceptions of African inferiority and subordination within global knowledge hierarchies.

Metrics such as Citation Impact (CI) and Category Normalized Citation Impact (CNCI), often treated as neutral indicators of research quality, are embedded within this same system of validation. While these metrics appear objective, they are calibrated against baselines derived largely from Global North publication ecosystems.  As a result, they implicitly privilege journals that are already indexed within dominant databases such as Scopus and Web of Science, where African journals remain underrepresented \citep{asubiaro}. In this sense, metrics do not simply measure impact; they participate in defining it. The use of global baselines in indicators such as CNCI further illustrates how epistemic power is standardised. By normalising citation performance against a global average, these metrics position African research within a framework shaped by external research priorities and citation practices. What is presented as expected impact is therefore not neutral, but historically and geographically situated.

In this study, the substantial differences in Category Normalized Citation Impact (CNCI) between African-based and non-African publication venues illustrate how these systems of validation are structured in ways that privilege certain publication spaces over others. As a result, citation impact, treated as a neutral indicator of research quality \citep{Boshoff_2018}  may instead reflect embedded hierarchies within global knowledge production.  Viewed through a decolonising lens, these patterns point to deeper structural inequities in how knowledge is valued, circulated, and legitimised.

Decolonising knowledge, in this context, requires more than increasing representation within existing systems. It entails a critical re-examination of the assumptions and infrastructures that determine where knowledge resides and how it becomes authoritative. As \cite{behari} cautions, dominant practices are often sustained through everyday academic behaviors and dispositions, making it necessary to remain attentive to how systems of validation are reproduced through routine scholarly activity.

\subsection{Knowledge Dependency and the Reproduction of Inequality}
Global knowledge systems anchor epistemic authority in particular locations, thereby creating recurring patterns of dependence that influence how knowledge is produced, disseminated, and valued over time. In the context of African astronomy, this dependency is not incidental but structurally reproduced through the very mechanisms that govern academic recognition and advancement. 

A central feature of this dynamic can be described as a validation trap. When professional recognition, promotion, and funding are linked to publication in externally validated journals, the value of African scholarship becomes dependent on approval from outside the continent \citep{asubiaro,ndlovu, un2023_knowledge_sovereignty}.  In this way, the international masthead functions not only as a quality marker, but as a gatekeeping mechanism through which epistemic legitimacy is conferred. Editorial boards, reviewers, and publishers located elsewhere therefore play a decisive role in determining which research questions are recognised as significant and which forms of knowledge are rendered visible. This dynamic fosters a cycle of academic dependency in which research agendas become increasingly oriented toward external expectations. Rather than emerging primarily from locally defined scientific priorities, knowledge production is shaped by the perceived demands of dominant publication systems. Scholars like \cite{crawford} and \cite{mungwini} challenge the idea that African knowledge can only gain legitimacy through validation from external, often Northern, institutions. They argue that this reliance on outside endorsement marginalises local knowledge and that African scholarship should be recognised on its own terms rather than through the lens of external approval.

Ownership and commercial control further reinforce this dependency. When research is published in international journals, copyright and intellectual property are often transferred from the author to commercial publishers. This results in a paradoxical situation in which African knowledge, frequently generated through locally situated observation and labour, is stored, distributed, and monetised through infrastructures located elsewhere. The financial model reinforces this imbalance: institutions may pay article processing charges to publish research and then pay again through subscription fees to access it \citep{onaolapo}. As a result, access to knowledge produced on the continent can become restricted to those with the financial means to engage with global publishing systems.
 
 \cite{un2023_knowledge_sovereignty} argues that Africa’s dependence on academic knowledge, policy guidance, and best-practice lessons from the West has undermined knowledge sovereignty, reinforcing epistemic injustices. This dependence erodes local scholarly infrastructure.  When high-status publications are concentrated in external venues, local journals and university presses struggle to attract quality submissions, which reduces their visibility, indexing, and perceived legitimacy. A self-reinforcing cycle emerges: fewer submissions lower impact, discouraging future contributions, and diminishing institutional support. The result is not just weakened local platforms, but a system that positions Africa primarily as a consumer rather than a sovereign producer of scientific knowledge \citep{crawford}.
This cycle can be seen as a form of intellectual displacement. Even when researchers remain physically based in Africa, their contributions are often absorbed into external systems of validation and circulation. The result is a persistent gap between where knowledge is produced and where it is recognised as authoritative.  \cite{mungwini} and \cite{crawford} argue, addressing epistemic injustice requires African institutions and scholars to reclaim and actively shape their own knowledge agendas, rather than remaining peripheral participants.

The empirical record developed in this paper, from manuscript astronomy in Timbuktu to contemporary bibliometric patterns, points to a central paradox: Africa is increasingly a site of astronomical observation and scientific labour, yet the durable record of that knowledge is frequently stored, validated, and valorised through infrastructures located elsewhere. 

\subsection{Toward Knowledge Sovereignty in African Astronomy}
In this study, knowledge sovereignty is understood not as isolation from global scholarship, but as the capacity to host, govern, and sustain the infrastructures through which knowledge becomes durable, discoverable, and legitimate.

The first pathway of this reconfiguration concerns material sovereignty, particularly in relation to data, archives, and computational infrastructure. In modern astronomy, epistemic authority is closely tied to control over data and the systems through which it is processed and accessed \citep{SKAOAccessRules2024}. While Africa is increasingly becoming a site of astronomical observation, this does not automatically translate into epistemic power \citep{Abebe_2021}.  Strengthening African participation in, and governance of, data repositories, science archives, and computational platforms is therefore central to ensuring that knowledge produced on the continent remains anchored within it. The main point here is this: if Africa hosts the telescopes but not the archives and workflows, authority will follow the archive rather than the sky.

A second pathway involves institutional sovereignty, particularly in relation to publication systems, rights, and visibility. The findings of this study highlight the extent to which scholarly validation is mediated by publication location and indexing structures. Developing African-governed publishing platforms, including equitable open access models, offers one way to address these asymmetries \citep{UNESCODiamondOA2024,DiamondOAActionPlan2022}. At the same time, strategies such as rights retention and repository development can ensure that research outputs remain accessible within African institutional ecosystems, even when published in international venues \citep{CoalitionSRightsRetention}. Attention to metadata, indexing, and discoverability is also critical, as visibility within global knowledge systems is often determined by technical infrastructures that are frequently treated as neutral but are, in practice, unevenly distributed \citep{Mongeon2016Coverage,Asubiaro2024RegionalDisparities,asubiaro}. In other words, if Africa exports prestige by default, it imports legitimacy at a high premium.

A third pathway is epistemic sovereignty, which concerns the authority to define research agendas, interpretive frameworks, and criteria of value. Metrics such as CI and CNCI, while useful for describing patterns of visibility, can reinforce dependency when treated as primary indicators of quality. Reforming research assessment practices to prioritise qualitative judgement, locally relevant contributions, and diverse forms of scholarly output is therefore essential \citep{DORA2013,CoARAAssessment2022,Hicks2015Leiden,Wilsdon2015MetricTide}. This includes recognising the value of data curation, software development, and infrastructure building, as well as research that responds to African-defined scientific and societal priorities.
Epistemic sovereignty also extends to curriculum and intellectual memory. The historical record demonstrates that Africa has long-standing traditions of astronomical knowledge and observation. Integrating these traditions into contemporary teaching and research does not require a rejection of global science, but rather a repositioning of African knowledge within it. In this sense, decolonisation is not simply a matter of inclusion, but of re-situating knowledge within a broader and more historically grounded epistemic landscape. As \cite{Chirikure2016} argues, meaningful transformation in higher education cannot begin with curriculum reform alone, but must start with the production and validation of knowledge itself. In this sense, epistemic sovereignty in African astronomy requires not only participation in global science, but the capacity to define what constitutes legitimate knowledge within the field.

\section{Limitations and Recommendations}\label{limit}
In this section, we will highlight some of the limitations of our study and outline some of our recommendations.
\subsection{Limitations of the Study}
This study relies on bibliometric data derived from the Web of Science Core Collection and associated InCites analytics. While these databases provide structured and widely used indicators of research output and citation impact, they are not neutral representations of global knowledge production. A well-documented limitation is the uneven coverage of journals across regions, with African-based journals significantly underrepresented relative to those in Europe and North America \citep{Mongeon2016Coverage,Asubiaro2024RegionalDisparities}.

As a result, this analysis necessarily measures African scientific output through the very infrastructures it seeks to interrogate. Citation Impact (CI) and Category Normalised Citation Impact (CNCI) reflect visibility and recognition within indexed systems, not intrinsic scientific value. Consequently, the low representation and citation performance of Africa-based journals in this study should not be interpreted as evidence of lower research quality, but rather as an artefact of limited index inclusion and weaker integration into global citation networks.

A further methodological limitation concerns the operational definition of \emph{African affiliated} and \emph{African-led publications}. In this study, African affiliation is identified through institutional address metadata, and leadership is proxied through first authorship. While these are standard bibliometric practices, they do not fully capture the complexity of intellectual contribution within large, multi-institutional collaborations. In fields such as astronomy, where consortia-based authorship is common, significant contributions may not be reflected in authorship position alone.
\subsection{Recommendations for Future Research}

The findings of this study point to several avenues for further investigation.

First, similar analyses should be conducted across other scientific disciplines to assess whether the patterns observed here are specific to astronomy or reflect broader systemic dynamics within global knowledge production. Comparative studies across fields such as physics, engineering, health sciences, education, and the social sciences would provide a more comprehensive understanding of the extent and variability of these asymmetries.

Second, future work should incorporate alternative data sources and methodological approaches that extend beyond conventional citation indexes. This may include regional databases, institutional repositories, and discipline - specific archives, as well as qualitative approaches that examine authorship practices, editorial governance, and peer review processes. Such work would help to triangulate bibliometric findings and reduce dependence on a single validation infrastructure.

Third, there is a need for more detailed studies of collaboration structures, particularly within large-scale international projects. Understanding how roles are distributed, how leadership is exercised, and how credit is assigned within these collaborations would provide deeper insight into the mechanisms underlying the affiliation - leadership gap identified in this study.

Finally, and most importantly, this study highlights the need for increased critical awareness within African research communities regarding the geopolitics of knowledge production. Scholars, institutions, and policymakers should engage more explicitly with questions of publication strategy, data governance, and research assessment, recognising that these are not merely technical decisions but structural determinants of epistemic authority.

In this sense, the broader recommendation is not prescriptive but reflexive: to encourage disciplines, institutions, and regions to examine how knowledge systems operate, whose interests they serve, and how they might be reshaped to support more equitable forms of scientific participation and authority.

\section{Conclusion}\label{conc}

This paper set out to examine a deceptively simple but structurally significant question: where does African astronomical knowledge reside? By situating contemporary astronomy within a long history of African sky knowledge, and by analysing publication and citation patterns within the Web of Science ecosystem, we have shown that the answer is not merely geographic, but material, institutional and epistemic.

Three main findings emerge. First, African astronomy is no longer peripheral to global scientific production. The continent has become an active and visible contributor to astronomical research, with African-affiliated publications achieving citation impacts above the global baseline. This reflects the success of major investments in infrastructure, human capital, and international collaboration.

Second, this participation does not translate into proportional epistemic authority. When leadership is proxied through first authorship, African-led outputs show substantially lower citation visibility. This affiliation-leadership gap indicates that while African researchers are integrated into global collaborations, agenda-setting, synthesis, and interpretive authority remain unevenly distributed.

Third, the location of publication and validation infrastructures plays a decisive role in shaping where knowledge becomes durable and authoritative. Only a negligible fraction of African-affiliated research is published in Africa-based journals, and these outputs exhibit markedly lower citation impact within indexed systems. This pattern does not reflect a deficit of knowledge production, but rather the structural marginalisation of African-controlled publication platforms within global indexing and evaluation regimes.

Taken together, these findings point to a broader conclusion: in contemporary astronomy, epistemic authority is inseparable from the infrastructures through which knowledge is stored, validated, and circulated. Africa participates in discovery, but the durable archive of that discovery is largely constituted elsewhere.

In this sense, African astronomy has become infrastructural; the remaining challenge is epistemic. Addressing this challenge requires not only investment in telescopes, data systems, and human capacity, but also a critical awareness of how global knowledge systems operate and how Africa is positioned within them. Without such awareness, well-intentioned participation can inadvertently reproduce the very structures that externalise African knowledge, concentrating its validation and ownership elsewhere.

The question, therefore, is not only how Africa contributes to global astronomy, but how it understands and navigates the systems through which knowledge acquires authority. This requires a deliberate shift from participation to residence. Knowledge sovereignty, as argued in this paper, is not about disengaging from global science, but about building and governing the material, institutional, and epistemic infrastructures that allow African astronomy to accumulate authority on its own terms.

This includes strengthening African-based publication systems, ensuring repository and data sovereignty, reforming research assessment practices, and embedding African intellectual traditions within the broader scientific narrative. Crucially, it also requires recognising Africa as a legitimate epistemic centre, capable of defining, interpreting, and sustaining knowledge about the universe.

As \cite{ndlovu} reminds us, African scholars must not remain “outsiders in our own land”. A return to the base, in this sense, entails taking seriously the location from which knowledge is produced, and affirming Africa as a site not only of observation, but of interpretation and authority.

Ultimately, the question is not whether Africa is part of global astronomy. It is whether Africa will also become a site where astronomical knowledge is curated, legitimised, and sustained. Without such a shift, the continent risks remaining, in epistemic terms, a contributor without custody.

\bibliographystyle{apalike}
\bibliography{references}

@incollection{Matorevhu2024,
  author    = {Matorevhu, Alois},
  title     = {Bibliometrics: Application Opportunities and Limitations},
  booktitle = {Bibliometrics - An Essential Methodological Tool for Research Projects},
  editor    = {de Oliveira, Ot{\'a}vio Jos{\'e}},
  publisher = {IntechOpen},
  year      = {2024},
  chapter   = {3},
  doi       = {10.5772/intechopen.1005292},
  url       = {https://www.intechopen.com/chapters/1181108},
  isbn      = {978-0-85466-802-1}
}

@article{Malville1998,
  author = {Malville, J. M. and Wendorf, F. and Mazar, A. and Schild, R.},
  title = {{Megaliths and Neolithic astronomy in southern Egypt}},
  journal = {Nature},
  volume = {392},
  pages = {488--491},
  year = {1998}
}

@incollection{Malville2003,
  author    = {Malville, J. McKim},
  title     = {{Astronomy at Nabta Playa}},
  booktitle = {Exploring the Ancient Skies: An Encyclopedic Survey of Archaeoastronomy},
  editor    = {Kelley, David H. and Milone, Eugene F.},
  publisher = {Springer},
  address   = {New York},
  year      = {2003}
}

@book{Wendorf2001,
  author = {Wendorf, F. and Schild, R.},
  title = {Holocene Settlement of the Egyptian Sahara. Volume 1: The Archaeology of Nabta Playa},
  publisher = {Kluwer Academic/Plenum},
  year = {2001}
}

@book{Hunwick2003,
  author = {Hunwick, John O.},
  title = {Timbuktu and the Songhay Empire},
  publisher = {Brill},
  year = {2003}
}

@book{Jeppie2008,
  editor = {Jeppie, Shamil and Diagne, Souleymane Bachir},
  title = {The Meanings of Timbuktu},
  publisher = {HSRC Press},
  year = {2008}
}

@incollection{Medupe2008,
  author = {Medupe, R. T.},
  title = {{Astronomy and the Timbuktu manuscripts}},
  booktitle = {The Meanings of Timbuktu},
  editor = {Jeppie, S. and Diagne, S. B.},
  publisher = {HSRC Press},
  year = {2008}
}

@book{Bleek1911,
  author    = {Bleek, Wilhelm H. I. and Lloyd, Lucy C.},
  title     = {Specimens of Bushman Folklore},
  publisher = {George Allen},
  address   = {London},
  year      = {1911}
}

@phdthesis{Hollmann2004,
  author = {Hollmann, Jeremy C.},
  title  = {The Queen of the Night: /Xam Oral Narratives and the Cosmos},
  school = {University of Cape Town},
  year   = {2004}
}

@book{Berglund1976,
  author    = {Berglund, Axel-Ivar},
  title     = {Zulu Thought-Patterns and Symbolism},
  publisher = {Indiana University Press},
  year      = {1976}
}

@book{Krige1950,
  author    = {Krige, Eileen Jensen},
  title     = {The Social System of the Zulus},
  publisher = {Shuter and Shooter},
  year      = {1950}
}

@book{Ruggles2015,
  author    = {Ruggles, Clive L. N.},
  title     = {Handbook of Archaeoastronomy and Ethnoastronomy},
  publisher = {Springer},
  year      = {2015}
}

@book{Saliba1994,
  author    = {Saliba, George},
  title     = {A History of Arabic Astronomy: Planetary Theories During the Golden Age of Islam},
  publisher = {New York University Press},
  year      = {1994}
}

@article{Brentjes2007,
  author  = {Brentjes, Sonja},
  title   = {{Islamic Astronomical Tradition in Sub-Saharan Africa}},
  journal = {Journal for the History of Arabic Science},
  volume  = {14},
  pages   = {1--32},
  year    = {2007}
}

@book{King2005,
  author    = {King, David A.},
  title     = {In Synchrony with the Heavens: Studies in Astronomical Timekeeping and Instrumentation in Medieval Islamic Civilization},
  publisher = {Brill},
  year      = {2005}
}

@book{Clagett1995,
  author    = {Clagett, Marshall},
  title     = {Ancient Egyptian Science, Volume II: Calendars, Clocks, and Astronomy},
  publisher = {American Philosophical Society},
  address   = {Philadelphia},
  year      = {1995}
}

@book{Neugebauer1969full,
  author    = {Neugebauer, Otto and Parker, Richard A.},
  title     = {Egyptian Astronomical Texts},
  publisher = {Brown University Press},
  address   = {Providence},
  year      = {1960--1969}
}

@book{Mosshammer2008,
  author    = {Mosshammer, Alden A.},
  title     = {The Easter Computus and the Origins of the Christian Era},
  publisher = {Oxford University Press},
  year      = {2008}
}

@article{Yaqob2007,
  author  = {Yaqob, Ayele},
  title   = {The {E}thiopian Calendar and the Computation of {E}aster},
  journal = {Journal of Ethiopian Studies},
  volume  = {40},
  pages   = {1--18},
  year    = {2007}
}

@book{Tamrat1972,
  author    = {Tamrat, Taddesse},
  title     = {Church and State in Ethiopia, 1270–1527},
  publisher = {Oxford University Press},
  year      = {1972}
}

@book{Haile1981,
  author    = {Haile, Getatchew},
  title     = {A Catalogue of {E}thiopian Manuscripts Microfilmed for the {E}thiopian Manuscript Microfilm Library},
  publisher = {Hill Monastic Manuscript Library},
  year      = {1981}
}

@book{Neugebauer1975,
  author    = {Neugebauer, Otto},
  title     = {A History of Ancient Mathematical Astronomy},
  publisher = {Springer},
  address   = {Berlin},
  year      = {1975}
}

@article{mignolo2002,
   author = {Mignolo, Walter D.},
   title = {{The Geopolitics of Knowledge and the Colonial Difference}},
   journal = {South Atlantic Quarterly},
   volume = {101},
   number = {1},
   pages = {57-96},
   ISSN = {0038-2876; 1527-8026; 0038-2876; 1527-8026},
   DOI = {10.1215/00382876-101-1-57},
   url = {http://dx.doi.org/10.1215/00382876-101-1-57},
   year = {2002},
   type = {Journal Article}
}

@article{mignolo2003,
   author = {Mignolo, Walter D.},
   title = {Globalization and the Geopolitics of Knowledge},
   journal = {Nepantla: Views from South},
   volume = {4},
   number = {1},
   pages = {97},
   ISSN = {1527-0858; 1527-0858},
   year = {2003},
   type = {Journal Article}
}

@article{ndlovu,
   author = {Ndlovu-Gatsheni, Sabelo},
   title = {THE DYNAMICS OF EPISTEMOLOGICAL DECOLONISATION IN THE 21ST CENTURY: TOWARDS EPISTEMIC FREEDOM},
   journal = {{The Strategic Review for Southern Africa}},
   volume = {40},
   number = {1},
   ISSN = {1013-1108; 1013-1108},
   DOI = {10.35293/srsa.v40i1.268},
   url = {http://dx.doi.org/10.35293/srsa.v40i1.268},
   year = {2020},
   type = {Journal Article}
}

@article{cloete,
   author = {Cloete, F.},
   title = {Moving Beyond Empty Decolonisation Mantras to Real Sustainable Empowerment in Africa},
   journal = {{Administratio Publica}},
   volume = {26},
   number = {1},
   pages = {58-85},
   ISSN = {1015-4833},
   DOI = {10.10520/ejc-adminpub-v26-n1-a4},
   year = {2018},
   type = {Journal Article}
}

@article{salgado,
   author = {Gandarilla Salgado, José Guadalupe and García-Bravo, María Haydeé and Benzi, Daniele},
   title = {{Two Decades of Aníbal Quijano’s Coloniality of Power, Eurocentrism and Latin America}},
   journal = {Contexto Internacional},
   volume = {43},
   number = {1},
   pages = {199-222},
   ISSN = {0102-8529; 0102-8529},
   year = {2021},
   type = {Journal Article}
}

@article{mungwini,
   author = {Mungwini, Pascah},
   title = {{“African Know Thyself”: Epistemic Injustice and the Quest for Liberative Knowledge}},
   journal = {International Journal of African Renaissance Studies - Multi-, Inter- and Transdisciplinarity},
   volume = {12},
   number = {2},
   pages = {5-18},
   ISSN = {1818-6874; 1753-7274; 1818-6874; 1753-7274},
   DOI = {10.1080/18186874.2017.1392125},
   url = {https://dx.doi.org/10.1080/18186874.2017.1392125},
   year = {2017},
   type = {Journal Article}
}

@misc{InCitesCitationImpact2024,
  author       = {{Clarivate}},
  title        = {Citation Impact},
  howpublished = {InCites Help Center (Indicators Handbook)},
  year         = {2024},
  note         = {Updated 29 April 2024},
  url          = {https://incites.zendesk.com/hc/en-gb/articles/24647918076177-Citation-Impact},
  urldate      = {2026-03-12}
}

@misc{InCitesCNCI2025,
  author       = {{Clarivate}},
  title        = {{Category Normalized Citation Impact (CNCI)}},
  howpublished = {InCites Help Center (Indicators Handbook)},
  year         = {2025},
  note         = {Updated 19 May 2025},
  url          = {https://incites.zendesk.com/hc/en-gb/articles/25087312115601-Category-Normalized-Citation-Impact-CNCI},
  urldate      = {2026-03-12}
}

@article{Waltman2016Review,
  author  = {Waltman, Ludo},
  title   = {A review of the literature on citation impact indicators},
  journal = {Journal of Informetrics},
  volume  = {10},
  number  = {2},
  pages   = {365--391},
  year    = {2016},
  doi     = {10.1016/j.joi.2016.02.007}
}

@article{Mongeon2016Coverage,
  author  = {Mongeon, Philippe and Paul-Hus, Ad{\`e}le},
  title   = {The journal coverage of {W}eb of {S}cience and {S}copus: a comparative analysis},
  journal = {Scientometrics},
  volume  = {106},
  number  = {1},
  pages   = {213--228},
  year    = {2016},
  doi     = {10.1007/s11192-015-1765-5}
}

@article{Asubiaro2024RegionalDisparities,
  author  = {Asubiaro, Toluwase Victor and Onaolapo, Sodiq and Mills, David S.},
  title   = {Regional disparities in {W}eb of {S}cience and {S}copus journal coverage},
  journal = {Scientometrics},
  volume  = {129},
  number  = {3},
  pages   = {1469--1491},
  year    = {2024},
  doi     = {10.1007/s11192-024-04948-x}
}

@article{Seglen1992JIF,
  author  = {Seglen, Per O.},
  title   = {How representative is the journal impact factor?},
  journal = {Research Evaluation},
  volume  = {2},
  number  = {3},
  pages   = {143--149},
  year    = {1992},
  doi     = {10.1093/rev/2.3.143}
}

@misc{DORA2013,
  author       = {{DORA}},
  title        = {{San Francisco Declaration on Research Assessment}},
  year         = {2013},
  url          = {https://sfdora.org/read/},
  note         = {Accessed: 2026-03-12}
}

@article{Hicks2015Leiden,
  author  = {Hicks, Diana and Wouters, Paul and Waltman, Ludo and de Rijcke, Sarah and Rafols, Ismael},
  title   = {Bibliometrics: The {Leiden Manifesto} for research metrics},
  journal = {Nature},
  volume  = {520},
  number  = {7548},
  pages   = {429--431},
  year    = {2015},
  doi     = {10.1038/520429a}
}

@techreport{Wilsdon2015MetricTide,
  author      = {Wilsdon and others},
  title       = {{The Metric Tide: Report of the Independent Review of the Role of Metrics in Research Assessment and Management}},
  institution = {Higher Education Funding Council for England (HEFCE)},
  year        = {2015},
  url         = {https://www.ukri.org/wp-content/uploads/2021/12/RE-151221-TheMetricTideFullReport2015.pdf},
  urldate     = {2026-03-12}
}

@misc{UNESCODiamondOA2024,
  author       = {{UNESCO}},
  title        = {{Diamond Open Access}},
  year         = {2024},
  url          = {https://www.unesco.org/en/diamond-open-access},
  urldate      = {2026-03-15}
}

@misc{CoARAAssessment2022,
  author       = {{CoARA}},
  title        = {Agreement on Reforming Research Assessment},
  year         = {2022},
  url          = {https://www.coara.org/wp-content/uploads/2025/11/2022_07_19_rra_agreement_final.pdf-3.pdf},
  urldate      = {2026-03-15},
  note         = {Signed version dated 20 July 2022}
}

@misc{SKAOAccessRules2024,
  author       = {{SKA Observatory (SKAO)}},
  title        = {{Access Rules and Regulations for the SKA Observatory}},
  year         = {2024},
  url          = {https://www.skao.int/sites/default/files/documents/SKAO-GOV-0000127-01_AccessRulesRegulations_Rev%2001%20-%20signed.pdf},
  urldate      = {2026-03-15}
}

@misc{CoalitionSRightsRetention,
  author       = {{cOAlition S}},
  title        = {Rights Retention Strategy},
  year         = {2021},
  howpublished = {\url{https://www.coalition-s.org/rights-retention-strategy/}},
  note         = {Accessed: 17 March 2026}
}

@misc{DiamondOAActionPlan2022,
  author       = {{Science Europe} and {cOAlition S} and {OPERAS} and {ANR}},
  title        = {{Action Plan for Diamond Open Access}},
  year         = {2022},
  url          = {https://www.scienceeurope.org/media/t3jgyo3u/202203-diamond-oa-action-plan.pdf},
  urldate      = {2026-03-15}
}

@article{asubiaro,
   author = {Asubiaro, Toluwase Victor and Onaolapo, Sodiq},
   title = {A comparative study of the coverage of African journals in Web of Science, Scopus, and CrossRef},
   journal = {Journal of the Association for Information Science and Technology},
   volume = {74},
   number = {7},
   pages = {745-758},
   ISSN = {2330-1635},
   DOI = {https://doi.org/10.1002/asi.24758},
   url = {https://asistdl.onlinelibrary.wiley.com/doi/abs/10.1002/asi.24758},
   year = {2023},
   type = {Journal Article}
}

@article{gebremariam,
   author = {Gebremariam, Eyob Balcha},
   title = {Decentering Coloniality: Epistemic Justice, Development Studies and Structural Transformation},
   journal = {The European Journal of Development Research},
   volume = {37},
   number = {2},
   pages = {442-453},
   ISSN = {1743-9728},
   DOI = {10.1057/s41287-024-00681-6},
   url = {https://doi.org/10.1057/s41287-024-00681-6},
   year = {2025},
   type = {Journal Article}
}

@article{Boshoff_2018,
 title={{Chris Callaghan’s criticism of the National Research Foundation’s rating methodology: A rebuttal}}, 
 volume={114}, 
 url={https://sajs.co.za/article/view/5142}, 
 DOI={10.17159/sajs.2018/a0278},
  number={7/8}, 
  journal={South African Journal of Science}, 
  author={Boshoff, Christo}, 
  year={2018}, month={Jul.} }

@article{behari,
   author = {Behari-Leak, K and Nkoala, S and Mokou, G and Binkowski, H},
   title = {Exploring disruptions of the coloniality of knowledge, power and being to enable agency as disciplinary activists for curriculum change},
   journal = {Journal of Decolonising Disciplines},
   volume = {2},
   number = {2},
   pages = {1-31},
   ISSN = {2664-3405},
   DOI = {10.35293/jdd.v2i2.25},
   url = {https://upjournals.up.ac.za/index.php/jdd/article/download/25/3398},
   year = {2020},
   type = {Journal Article}
}

@misc{un2023_knowledge_sovereignty,
  author       = {Gebremariam, Eyob Balcha},
  title        = {Restoring {Africa’s Knowledge Sovereignty Key to Sustainable Development}},
  journal  =  {Office of the Special Adviser on Africa},
  year         = {2023},
  url          = {https://www.un.org/osaa/news/restoring-africa%E2%80%99s-knowledge-sovereignty-key-sustainable-development},
  note         = {Accessed: 2026-03-18}
}

@article{crawford,
   author = {Crawford, Gordon and Mai-Bornu, Zainab and Landstrm, Karl},
   title = {{Decolonising knowledge production on Africa}: why its still necessary and what can be done},
   journal = {Journal of the British Academy},
   volume = {9s1},
   pages = {21-46},
   ISSN = {2052-7217; 2052-7217},
   DOI = {10.5871/jba/009s1.021},
   url = {http://dx.doi.org/10.5871/jba/009s1.021},
   year = {2021},
   type = {Journal Article}
}

@misc{Chirikure2016,
  author    = {Chirikure, Shadreck},
  title     = {Transforming higher education: first comes knowledge, then curriculum},
  year      = {2016},
  url       = {https://theconversation.com/transforming-higher-education-first-comes-knowledge-then-curriculum-64833},
  urldate   = {2026-03-24},
  publisher = {The Conversation}
}

@inproceedings{Abebe_2021, series={FAccT ’21},
   title={Narratives and Counternarratives on Data Sharing in Africa},
   url={http://dx.doi.org/10.1145/3442188.3445897},
   DOI={10.1145/3442188.3445897},
   booktitle={Proceedings of the 2021 ACM Conference on Fairness, Accountability, and Transparency},
   publisher={ACM},
   author={Abebe, Rediet and Aruleba, Kehinde and Birhane, Abeba and Kingsley, Sara and Obaido, George and Remy, Sekou L. and Sadagopan, Swathi},
   year={2021},
   month=mar, pages={329–341},
   collection={FAccT ’21} }

@article{onaolapo,
author = {Onaolapo, Sodiq and Ayeni, Philips and Mncube, Siphamandla},
year = {2025},
month = {06},
pages = {},
title = {Open access publishing in an African context: Notable improvements and recurring challenges},
journal = {IFLA Journal},
doi = {10.1177/03400352251351113}
}

\newpage

\appendix
\renewcommand{\thetable}{\Alph{table}}
\renewcommand{\tablename}{Appendix Table}
\setcounter{table}{0}

\clearpage
\begin{table}[!t]
\centering
\caption{Africa: who are we collaborating with? Publication and citation metrics by collaborating country.}
\label{tab:africa_collaboration_countries}

\setlength{\tabcolsep}{3pt}
\renewcommand{\arraystretch}{0.9}

{\tiny
\begin{minipage}[t]{0.48\textwidth}
\centering
\begin{tabular}{>{\RaggedRight\arraybackslash}p{3.2cm}rrrr}
\toprule
\textbf{Name} & \textbf{Docs} & \textbf{Times cited} & \textbf{CI} & \textbf{CNCI} \\
\midrule
Baseline for all items & 12846 & 555743 & 43.26 & 1.76 \\
South Africa & 9489 & 462687 & 48.76 & 1.82 \\
USA & 6543 & 416957 & 63.73 & 2.38 \\
United Kingdom & 5709 & 390573 & 68.41 & 2.55 \\
England & 5392 & 379829 & 70.44 & 2.61 \\
Germany & 4773 & 348036 & 72.92 & 2.70 \\
France & 4151 & 308954 & 74.43 & 2.83 \\
Italy & 3886 & 267472 & 68.83 & 2.78 \\
Spain & 3170 & 257408 & 81.20 & 3.12 \\
Australia & 3125 & 221506 & 70.88 & 2.64 \\
Netherlands & 2630 & 223901 & 85.13 & 3.40 \\
China mainland & 2526 & 165421 & 65.49 & 2.76 \\
Poland & 2296 & 193658 & 84.35 & 2.99 \\
India & 2281 & 161654 & 70.87 & 2.69 \\
Chile & 2196 & 185862 & 84.64 & 3.36 \\
Canada & 2117 & 192916 & 91.13 & 3.48 \\
Switzerland & 2050 & 204214 & 99.62 & 3.70 \\
Japan & 2001 & 149996 & 74.96 & 2.99 \\
Russia & 1904 & 154572 & 81.18 & 2.67 \\
Brazil & 1889 & 141365 & 74.84 & 2.83 \\
Austria & 1648 & 120594 & 73.18 & 2.65 \\
Sweden & 1572 & 157408 & 100.13 & 4.04 \\
Egypt & 1557 & 43707 & 28.07 & 0.96 \\
Czech Republic & 1524 & 110524 & 72.52 & 2.53 \\
Scotland & 1479 & 125194 & 84.65 & 3.42 \\
Greece & 1414 & 102240 & 72.31 & 2.93 \\
Portugal & 1346 & 109082 & 81.04 & 3.02 \\
Taiwan & 1334 & 116489 & 87.32 & 3.14 \\
Denmark & 1302 & 151328 & 116.23 & 4.59 \\
Turkiye & 1301 & 77672 & 59.70 & 2.00 \\
Finland & 1296 & 142138 & 109.67 & 4.31 \\
Hungary & 1273 & 106697 & 83.82 & 3.08 \\
Belgium & 1256 & 90063 & 71.71 & 2.94 \\
Armenia & 1238 & 83529 & 67.47 & 1.99 \\
Mexico & 1199 & 88584 & 73.88 & 2.81 \\
South Korea & 1193 & 85194 & 71.41 & 2.71 \\
Romania & 1043 & 85937 & 82.39 & 2.79 \\
Serbia & 1016 & 62103 & 61.12 & 2.01 \\
Norway & 1010 & 99463 & 98.48 & 3.65 \\
Ireland & 965 & 93301 & 96.68 & 3.52 \\
Colombia & 952 & 58467 & 61.41 & 1.98 \\
Morocco & 951 & 45595 & 47.94 & 1.78 \\
Georgia & 940 & 57560 & 61.23 & 1.94 \\
Israel & 834 & 74013 & 88.74 & 3.70 \\
Croatia & 834 & 54891 & 65.82 & 2.64 \\
Ukraine & 825 & 43597 & 52.84 & 1.82 \\
New Zealand & 823 & 49202 & 59.78 & 1.86 \\
Bulgaria & 794 & 39133 & 49.29 & 1.78 \\
Belarus & 746 & 53043 & 71.10 & 1.98 \\
Thailand & 713 & 34870 & 48.91 & 1.89 \\
Slovakia & 680 & 44875 & 65.99 & 2.14 \\
Pakistan & 672 & 35031 & 52.13 & 1.81 \\
Saudi Arabia & 643 & 19330 & 30.06 & 1.23 \\
Iran & 612 & 48679 & 79.54 & 2.88 \\
Azerbaijan & 606 & 31419 & 51.85 & 1.86 \\
Argentina & 591 & 42259 & 71.50 & 2.75 \\
Slovenia & 581 & 63725 & 109.68 & 4.11 \\
Malaysia & 579 & 31598 & 54.57 & 2.35 \\
Cyprus & 515 & 27848 & 54.07 & 1.82 \\
Estonia & 496 & 32534 & 65.59 & 2.12 \\
Lithuania & 487 & 32052 & 65.82 & 2.08 \\
Wales & 450 & 75284 & 167.30 & 6.45 \\
Hong Kong & 414 & 30046 & 72.57 & 3.49 \\
Algeria & 411 & 28522 & 69.40 & 3.65 \\
Nigeria & 411 & 5812 & 14.14 & 0.73 \\
Qatar & 403 & 15491 & 38.44 & 1.38 \\
Sri Lanka & 355 & 14945 & 42.10 & 1.49 \\
United Arab Emirates & 324 & 10306 & 31.81 & 1.73 \\
Namibia & 280 & 22661 & 80.93 & 2.59 \\
Ecuador & 277 & 8524 & 30.77 & 1.36 \\
Latvia & 266 & 8484 & 31.89 & 1.38 \\
Ethiopia & 250 & 3739 & 14.96 & 0.75 \\
Northern Ireland & 206 & 19887 & 96.54 & 5.88 \\
Peru & 200 & 10280 & 51.40 & 1.74 \\
Uzbekistan & 190 & 6191 & 32.58 & 2.22 \\
Cuba & 180 & 9706 & 53.92 & 1.81 \\
Tunisia & 179 & 1940 & 10.84 & 0.45 \\
Palestine & 156 & 4126 & 26.45 & 1.27 \\
\bottomrule
\end{tabular}
\end{minipage}
\hfill
\begin{minipage}[t]{0.48\textwidth}
\centering
\begin{tabular}{>{\RaggedRight\arraybackslash}p{3.2cm}rrrr}
\toprule
\textbf{Name} & \textbf{Docs} & \textbf{Times cited} & \textbf{CI} & \textbf{CNCI} \\
\midrule
Montenegro & 139 & 2510 & 18.06 & 1.19 \\
Kuwait & 130 & 1847 & 14.21 & 1.78 \\
Indonesia & 128 & 3851 & 30.09 & 1.47 \\
Philippines & 108 & 1499 & 13.88 & 1.50 \\
Uganda & 95 & 1663 & 17.51 & 0.87 \\
Burkina Faso & 83 & 1519 & 18.30 & 1.24 \\
Kazakhstan & 80 & 4345 & 54.31 & 3.01 \\
Oman & 79 & 2906 & 36.78 & 2.41 \\
Benin & 77 & 2320 & 30.13 & 0.97 \\
Iceland & 70 & 13081 & 186.87 & 6.67 \\
Kenya & 63 & 991 & 15.73 & 0.97 \\
Lebanon & 60 & 6187 & 103.12 & 6.62 \\
Botswana & 60 & 585 & 9.75 & 0.66 \\
Malta & 55 & 4473 & 81.33 & 8.14 \\
Mongolia & 51 & 3039 & 59.59 & 2.33 \\
Madagascar & 50 & 1075 & 21.50 & 1.11 \\
Rwanda & 40 & 672 & 16.80 & 0.73 \\
Cameroon & 35 & 672 & 19.20 & 0.76 \\
Cote d'Ivoire & 33 & 746 & 22.61 & 0.85 \\
Tanzania & 33 & 345 & 10.45 & 0.68 \\
Sudan & 32 & 701 & 21.91 & 0.95 \\
Ghana & 32 & 269 & 8.41 & 0.47 \\
Vatican & 31 & 3721 & 120.03 & 6.52 \\
Vietnam & 30 & 2504 & 83.47 & 2.93 \\
Senegal & 26 & 497 & 19.12 & 1.00 \\
Zambia & 21 & 204 & 9.71 & 0.80 \\
Bangladesh & 18 & 749 & 41.61 & 4.15 \\
Mauritius & 16 & 50 & 3.12 & 0.09 \\
Syria & 15 & 296 & 19.73 & 0.49 \\
Macedonia & 13 & 839 & 64.54 & 7.79 \\
Niger & 11 & 294 & 26.73 & 1.38 \\
Bahrain & 10 & 108 & 10.80 & 0.48 \\
Libya & 10 & 100 & 10.00 & 0.23 \\
Iraq & 10 & 48 & 4.80 & 0.72 \\
Eritrea & 8 & 196 & 24.50 & 0.51 \\
Honduras & 7 & 334 & 47.71 & 1.94 \\
Venezuela & 7 & 223 & 31.86 & 1.25 \\
Paraguay & 7 & 72 & 10.29 & 0.74 \\
Cape Verde & 6 & 91 & 15.17 & 0.49 \\
Singapore & 6 & 85 & 14.17 & 1.06 \\
Burundi & 6 & 56 & 9.33 & 0.38 \\
Congo Democratic Republic & 6 & 41 & 6.83 & 0.50 \\
Monaco & 5 & 3763 & 752.60 & 23.86 \\
Luxembourg & 5 & 137 & 27.40 & 8.27 \\
Jordan & 5 & 61 & 12.20 & 4.87 \\
Chad & 5 & 52 & 10.40 & 0.29 \\
Mozambique & 5 & 28 & 5.60 & 0.25 \\
Gambia & 5 & 28 & 5.60 & 1.12 \\
Nepal & 5 & 12 & 2.40 & 0.32 \\
Bosnia \& Herzegovina & 4 & 735 & 183.75 & 17.09 \\
Kyrgyzstan & 4 & 17 & 4.25 & 0.19 \\
Guinea & 4 & 6 & 1.50 & 0.05 \\
Albania & 3 & 470 & 156.67 & 25.33 \\
Yemen & 3 & 18 & 6.00 & 0.23 \\
Malawi & 3 & 9 & 3.00 & 0.42 \\
Lesotho & 2 & 124 & 62.00 & 1.28 \\
Fiji & 2 & 46 & 23.00 & 0.83 \\
Togo & 2 & 34 & 17.00 & 0.50 \\
Uruguay & 2 & 34 & 17.00 & 0.54 \\
Macau & 2 & 27 & 13.50 & 0.56 \\
Costa Rica & 2 & 14 & 7.00 & 1.07 \\
Moldova & 1 & 149 & 149.00 & 3.44 \\
Nicaragua & 1 & 80 & 80.00 & 1.72 \\
Bahamas & 1 & 78 & 78.00 & 36.54 \\
Zimbabwe & 1 & 65 & 65.00 & 1.31 \\
Barbados & 1 & 32 & 32.00 & 0.71 \\
Trinidad \& Tobago & 1 & 32 & 32.00 & 0.71 \\
Brunei & 1 & 13 & 13.00 & 2.03 \\
Central African Republic & 1 & 12 & 12.00 & 1.88 \\
Marshall Islands & 1 & 11 & 11.00 & 0.57 \\
Eswatini & 1 & 10 & 10.00 & 0.41 \\
Liberia & 1 & 2 & 2.00 & 0.18 \\
Saint Lucia & 1 & 1 & 1.00 & 5.61 \\
Bolivia & 1 & 0 & 0.00 & 0.00 \\
Angola & 1 & 0 & 0.00 & 0.00 \\
Comoros & 1 & 0 & 0.00 & 0.00 \\
\bottomrule
\end{tabular}
\end{minipage}
}
\end{table}

\clearpage
\begin{table*}[!t]
\centering
\small
\caption{Africa: publication and citation impact by country.}
\label{tab:africa_country_cnci}
\begin{tabular}{L{2.2cm}rrrr | L{2.2cm}rrrr}
\toprule
\textbf{Name} & \textbf{Docs} & \textbf{Cited} & \textbf{Rank} & \textbf{CNCI} &
\textbf{Name} & \textbf{Docs} & \textbf{Cited} & \textbf{Rank} & \textbf{CNCI} \\
\midrule

Global baseline & 647347 & 18586142 & -- & 1.06 & Niger & 11 & 294 & 22 & 1.38 \\
African baseline & 16023 & 596102 & -- & 1.52 & Zambia & 21 & 204 & 23 & 0.80 \\
South Africa & 11052 & 489114 & 1 & 1.66 & Eritrea & 8 & 196 & 24 & 0.51 \\
Egypt & 2176 & 49567 & 2 & 0.79 & Libya & 14 & 170 & 25 & 0.30 \\
Morocco & 1151 & 47287 & 3 & 1.57 & Lesotho & 2 & 124 & 26 & 1.28 \\
Algeria & 670 & 30595 & 4 & 2.38 & Cape Verde & 6 & 91 & 27 & 0.49 \\
Namibia & 283 & 22669 & 5 & 2.57 & Zimbabwe & 11 & 91 & 27 & 0.19 \\
Nigeria & 689 & 8425 & 6 & 0.57 & Mauritius & 27 & 61 & 29 & 0.06 \\
Ethiopia & 327 & 4087 & 7 & 0.62 & Burundi & 6 & 56 & 30 & 0.38 \\
Benin & 100 & 2638 & 8 & 0.87 & Chad & 5 & 52 & 31 & 0.29 \\
Tunisia & 225 & 2162 & 9 & 0.39 & Togo & 4 & 43 & 32 & 0.36 \\
Uganda & 108 & 1763 & 10 & 0.80 & DR Congo & 6 & 41 & 33 & 0.50 \\
Burkina Faso & 86 & 1526 & 11 & 1.21 & Mozambique & 5 & 28 & 34 & 0.25 \\
Madagascar & 50 & 1075 & 12 & 1.11 & Gambia & 5 & 28 & 34 & 1.12 \\
Cameroon & 71 & 1048 & 13 & 0.56 & CAR & 1 & 12 & 36 & 1.88 \\
Kenya & 69 & 1018 & 14 & 0.90 & Mali & 2 & 9 & 37 & 0.25 \\
Sudan & 49 & 842 & 15 & 0.68 & Malawi & 3 & 9 & 37 & 0.42 \\
Cote d'Ivoire & 36 & 787 & 16 & 0.81 & Guinea & 4 & 6 & 39 & 0.05 \\
Rwanda & 43 & 674 & 17 & 0.69 & Liberia & 1 & 2 & 40 & 0.18 \\
Botswana & 64 & 595 & 18 & 0.63 & Angola & 1 & 0 & 41 & 0.00 \\
Senegal & 29 & 510 & 19 & 0.92 & Comoros & 1 & 0 & 41 & 0.00 \\
Tanzania & 42 & 388 & 20 & 0.63 &  &  &  &  &  \\
Ghana & 37 & 371 & 21 & 0.53 &  &  &  &  &  \\

\bottomrule
\end{tabular}
\end{table*}

\begin{table}[p]
\centering
\caption{Publication and citation metrics by country/territory.}
\label{tab:wos_country_metrics}

\setlength{\tabcolsep}{3pt}
\renewcommand{\arraystretch}{0.9}

{\tiny
\begin{minipage}[t]{0.48\textwidth}\vspace{0pt}
\centering
\begin{tabular}{>{\RaggedRight\arraybackslash}p{3.25cm}rrrr}
\toprule
\textbf{Name} & \textbf{Docs} & \textbf{Times cited} & \textbf{CI} & \textbf{CNCI} \\
\midrule
Publications worldwide on WoS & 647347 & 18586142 & 28.71 & 1.06 \\
Publications worldwide on WoS with listed country & 633556 & 18527786 & 29.24 & 1.07 \\
USA & 254706 & 10707210 & 42.04 & 1.37 \\
Germany & 105894 & 4543832 & 42.91 & 1.45 \\
United Kingdom & 98331 & 4343129 & 44.17 & 1.48 \\
England & 86929 & 3960820 & 45.56 & 1.52 \\
France & 73279 & 3120856 & 42.59 & 1.48 \\
Italy & 71521 & 2687412 & 37.58 & 1.36 \\
China mainland & 67428 & 1562532 & 23.17 & 0.98 \\
Japan & 52875 & 1869568 & 35.36 & 1.23 \\
Spain & 50941 & 1989680 & 39.06 & 1.46 \\
Russia & 43649 & 1122914 & 25.73 & 0.81 \\
Canada & 36924 & 1787646 & 48.41 & 1.67 \\
Netherlands & 33490 & 1636190 & 48.86 & 1.78 \\
India & 32636 & 808958 & 24.79 & 0.97 \\
Australia & 30352 & 1375024 & 45.30 & 1.66 \\
Switzerland & 26162 & 1337623 & 51.13 & 1.88 \\
Chile & 24848 & 1032048 & 41.53 & 1.54 \\
Brazil & 21084 & 659059 & 31.26 & 1.12 \\
Poland & 19724 & 724857 & 36.75 & 1.28 \\
South Korea & 18135 & 578333 & 31.89 & 1.15 \\
Sweden & 17497 & 782795 & 44.74 & 1.61 \\
Belgium & 15890 & 627257 & 39.47 & 1.46 \\
Scotland & 15719 & 846854 & 53.87 & 1.86 \\
Mexico & 14984 & 449502 & 30.00 & 1.10 \\
Taiwan & 12312 & 470061 & 38.18 & 1.43 \\
Denmark & 12066 & 650571 & 53.92 & 2.15 \\
Austria & 11288 & 448531 & 39.74 & 1.42 \\
\textbf{South Africa} & 11052 & 489114 & 44.26 & 1.66 \\
Israel & 10709 & 526355 & 49.15 & 1.77 \\
Czech Republic & 10439 & 322168 & 30.86 & 1.21 \\
Finland & 10092 & 441956 & 43.79 & 1.54 \\
Portugal & 9302 & 432487 & 46.49 & 1.67 \\
Greece & 9238 & 331993 & 35.94 & 1.38 \\
Argentina & 8162 & 234993 & 28.79 & 0.97 \\
Ukraine & 7690 & 160905 & 20.92 & 0.67 \\
Hungary & 7144 & 335654 & 46.98 & 1.71 \\
Turkiye & 6559 & 172352 & 26.28 & 0.99 \\
Norway & 6126 & 277058 & 45.23 & 1.64 \\
Iran & 6037 & 142823 & 23.66 & 0.90 \\
Ireland & 5878 & 264025 & 44.92 & 1.76 \\
Wales & 4705 & 279174 & 59.34 & 2.17 \\
Northern Ireland & 3847 & 155211 & 40.35 & 1.54 \\
Romania & 3793 & 138016 & 36.39 & 1.35 \\
Bulgaria & 3412 & 103273 & 30.27 & 0.95 \\
New Zealand & 3405 & 132078 & 38.79 & 1.43 \\
Slovakia & 3067 & 82447 & 26.88 & 0.90 \\
Pakistan & 3004 & 74909 & 24.94 & 1.09 \\
Armenia & 2981 & 117806 & 39.52 & 1.15 \\
Colombia & 2771 & 96705 & 34.90 & 1.53 \\
Hong Kong & 2768 & 104740 & 37.84 & 1.70 \\
Serbia & 2726 & 87539 & 32.11 & 1.14 \\
Croatia & 2628 & 113984 & 43.37 & 1.59 \\
Thailand & 2452 & 65310 & 26.64 & 1.22 \\
Georgia & 2211 & 100089 & 45.27 & 1.64 \\
Slovenia & 2193 & 131610 & 60.01 & 2.22 \\
\textbf{Egypt} & 2176 & 49567 & 22.78 & 0.79 \\
Saudi Arabia & 1901 & 49461 & 26.02 & 0.99 \\
Estonia & 1794 & 71683 & 39.96 & 1.54 \\
United Arab Emirates & 1563 & 39674 & 25.38 & 1.36 \\
Kazakhstan & 1250 & 26413 & 21.13 & 1.06 \\
Cyprus & 1176 & 44538 & 37.87 & 1.44 \\
Malaysia & 1168 & 39424 & 33.75 & 1.55 \\
\textbf{Morocco} & 1151 & 47287 & 41.08 & 1.57 \\
Lithuania & 1150 & 44281 & 38.51 & 1.31 \\
Azerbaijan & 1132 & 35623 & 31.47 & 1.44 \\
Belarus & 1053 & 57456 & 54.56 & 1.53 \\
Uzbekistan & 1031 & 25428 & 24.66 & 1.33 \\
Venezuela & 826 & 42319 & 51.23 & 1.19 \\
Vietnam & 796 & 19491 & 24.49 & 0.93 \\
Ecuador & 729 & 19722 & 27.05 & 1.00 \\
\textbf{Nigeria} & 689 & 8425 & 12.23 & 0.57 \\
\textbf{Algeria} & 670 & 30595 & 45.66 & 2.38 \\
Indonesia & 643 & 9871 & 15.35 & 0.74 \\
Iceland & 609 & 41662 & 68.41 & 2.31 \\
Peru & 574 & 17041 & 29.69 & 1.07 \\
Macau & 531 & 8291 & 15.61 & 0.69 \\
Qatar & 507 & 17827 & 35.16 & 1.24 \\
Latvia & 493 & 10712 & 21.73 & 0.99 \\
Vatican & 455 & 19511 & 42.88 & 1.72 \\
Mongolia & 448 & 6924 & 15.46 & 0.69 \\
Sri Lanka & 409 & 15269 & 37.33 & 1.44 \\
Cuba & 382 & 13062 & 34.19 & 1.11 \\
Singapore & 372 & 6255 & 16.81 & 0.72 \\
Malta & 366 & 15121 & 41.31 & 3.57 \\
Lebanon & 336 & 11428 & 34.01 & 1.85 \\
\textbf{Ethiopia} & 327 & 4087 & 12.50 & 0.62 \\
Uruguay & 326 & 8296 & 25.45 & 1.02 \\
\textbf{Namibia} & 283 & 22669 & 80.10 & 2.57 \\
Philippines & 266 & 5093 & 19.15 & 1.34 \\
\bottomrule
\end{tabular}
\end{minipage}
\hfill
\begin{minipage}[t]{0.48\textwidth}\vspace{0pt}
\centering
\begin{tabular}{>{\RaggedRight\arraybackslash}p{3.25cm}rrrr}
\toprule
\textbf{Name} & \textbf{Docs} & \textbf{Times cited} & \textbf{CI} & \textbf{CNCI} \\
\midrule
\textbf{Tunisia} & 225 & 2162 & 9.61 & 0.39 \\
Bangladesh & 219 & 3361 & 15.35 & 0.91 \\
Oman & 213 & 4864 & 22.84 & 1.90 \\
Kuwait & 205 & 3700 & 18.05 & 1.75 \\
Palestine & 176 & 4459 & 25.34 & 1.20 \\
Jordan & 177 & 1921 & 10.85 & 0.96 \\
Montenegro & 175 & 3223 & 18.42 & 1.05 \\
Yugoslavia & 165 & 1447 & 8.77 & 0.22 \\
Costa Rica & 160 & 3340 & 20.88 & 0.69 \\
Iraq & 136 & 1747 & 12.85 & 0.65 \\
Macedonia & 118 & 4450 & 37.71 & 2.25 \\
\textbf{Uganda} & 108 & 1763 & 16.32 & 0.80 \\
Tajikistan & 107 & 938 & 8.77 & 0.27 \\
Nepal & 103 & 1082 & 10.50 & 0.56 \\
\textbf{Benin} & 100 & 2638 & 26.38 & 0.87 \\
Kyrgyzstan & 100 & 1006 & 10.06 & 0.31 \\
\textbf{Burkina Faso} & 86 & 1526 & 17.74 & 1.21 \\
Monaco & 83 & 17448 & 210.22 & 8.62 \\
Luxembourg & 81 & 1444 & 17.83 & 1.51 \\
\textbf{Cameroon} & 71 & 1048 & 14.76 & 0.56 \\
\textbf{Kenya} & 69 & 1018 & 14.75 & 0.90 \\
Honduras & 68 & 942 & 13.85 & 0.88 \\
\textbf{Botswana} & 64 & 595 & 9.30 & 0.63 \\
Bolivia & 56 & 2022 & 36.11 & 0.95 \\
Serbia \& Montenegro & 50 & 1607 & 32.14 & 0.67 \\
\textbf{Madagascar} & 50 & 1075 & 21.50 & 1.11 \\
\textbf{Sudan} & 49 & 842 & 17.18 & 0.68 \\
\textbf{Rwanda} & 43 & 674 & 15.67 & 0.69 \\
\textbf{Tanzania} & 42 & 388 & 9.24 & 0.63 \\
Bosnia \& Herzegovina & 40 & 1993 & 49.83 & 2.69 \\
Bahamas & 40 & 649 & 16.23 & 1.92 \\
Guatemala & 38 & 270 & 7.11 & 0.26 \\
\textbf{Ghana} & 37 & 371 & 10.03 & 0.53 \\
\textbf{Cote d'Ivoire} & 36 & 787 & 21.86 & 0.81 \\
Syria & 36 & 460 & 12.78 & 0.41 \\
Barbados & 35 & 2409 & 68.83 & 1.67 \\
Albania & 34 & 642 & 18.88 & 2.46 \\
Fiji & 30 & 685 & 22.83 & 0.65 \\
Trinidad \& Tobago & 29 & 929 & 32.03 & 0.81 \\
\textbf{Senegal} & 29 & 510 & 17.59 & 0.92 \\
Bahrain & 27 & 219 & 8.11 & 0.33 \\
\textbf{Mauritius} & 27 & 61 & 2.26 & 0.06 \\
\textbf{Zambia} & 21 & 204 & 9.71 & 0.80 \\
Paraguay & 20 & 489 & 24.45 & 1.05 \\
Yemen & 20 & 277 & 13.85 & 0.44 \\
Moldova & 18 & 614 & 34.11 & 0.96 \\
Panama & 15 & 392 & 26.13 & 0.60 \\
North Korea & 15 & 143 & 9.53 & 0.26 \\
\textbf{Libya} & 14 & 170 & 12.14 & 0.30 \\
Nicaragua & 12 & 142 & 11.83 & 0.41 \\
\textbf{Niger} & 11 & 294 & 26.73 & 1.38 \\
\textbf{Zimbabwe} & 11 & 91 & 8.27 & 0.19 \\
Afghanistan & 10 & 46 & 4.60 & 1.04 \\
Jamaica & 9 & 117 & 13.00 & 0.53 \\
\textbf{Eritrea} & 8 & 196 & 24.50 & 0.51 \\
Andorra & 7 & 362 & 51.71 & 1.37 \\
Grenada & 7 & 48 & 6.86 & 0.33 \\
El Salvador & 7 & 22 & 3.14 & 0.41 \\
\textbf{Cape Verde} & 6 & 91 & 15.17 & 0.49 \\
\textbf{Burundi} & 6 & 56 & 9.33 & 0.38 \\
\textbf{Congo Democratic Republic} & 6 & 41 & 6.83 & 0.50 \\
\textbf{Chad} & 5 & 52 & 10.40 & 0.29 \\
\textbf{Mozambique} & 5 & 28 & 5.60 & 0.25 \\
\textbf{Gambia} & 5 & 28 & 5.60 & 1.12 \\
Brunei & 5 & 26 & 5.20 & 0.46 \\
Haiti & 4 & 48 & 12.00 & 0.93 \\
\textbf{Togo} & 4 & 43 & 10.75 & 0.36 \\
Myanmar & 4 & 32 & 8.00 & 1.38 \\
\textbf{Guinea} & 4 & 6 & 1.50 & 0.05 \\
Marshall Islands & 3 & 87 & 29.00 & 0.71 \\
Liechtenstein & 3 & 28 & 9.33 & 0.19 \\
\textbf{Malawi} & 3 & 9 & 3.00 & 0.42 \\
Saint Kitts \& Nevis & 2 & 156 & 78.00 & 1.71 \\
\textbf{Lesotho} & 2 & 124 & 62.00 & 1.28 \\
Cook Islands & 2 & 50 & 25.00 & 1.09 \\
French Guiana & 2 & 19 & 9.50 & 0.20 \\
Laos & 2 & 16 & 8.00 & 0.35 \\
Vanuatu & 2 & 14 & 7.00 & 0.33 \\
\textbf{Eswatini} & 2 & 12 & 6.00 & 0.24 \\
\textbf{Mali} & 2 & 9 & 4.50 & 0.25 \\
Greenland & 1 & 29 & 29.00 & 0.93 \\
Cambodia & 1 & 24 & 24.00 & 0.88 \\
\textbf{Central African Republic} & 1 & 12 & 12.00 & 1.88 \\
New Caledonia & 1 & 9 & 9.00 & 0.21 \\
Dominican Republic & 1 & 4 & 4.00 & 0.16 \\
Guyana & 1 & 4 & 4.00 & 0.08 \\
Kiribati & 1 & 2 & 2.00 & 1.13 \\
\textbf{Liberia} & 1 & 2 & 2.00 & 0.18 \\
Saint Lucia & 1 & 1 & 1.00 & 5.61 \\
Papua New Guinea & 1 & 0 & 0.00 & 0.00 \\
\textbf{Angola} & 1 & 0 & 0.00 & 0.00 \\
\textbf{Comoros} & 1 & 0 & 0.00 & 0.00 \\
\bottomrule
\end{tabular}
\end{minipage}
}
\end{table}

\end{document}